\documentclass[journal,10pt]{IEEEtran}

\ifCLASSINFOpdf
 \usepackage[pdftex]{graphicx}
   \graphicspath{{../pdf/}{../jpeg/}}
	 \DeclareGraphicsExtensions{.pdf,.jpeg,.png}
\else
   \usepackage[dvips]{graphicx}
   \graphicspath{{./eps/}}
   \DeclareGraphicsExtensions{.eps}
\fi

\usepackage[cmex10]{amsmath}

\usepackage{array}

\ifCLASSOPTIONcompsoc
  \usepackage[caption=false,font=normalsize,labelfont=sf,textfont=sf]{subfig}
\else
  \usepackage[caption=false,font=footnotesize]{subfig}

\usepackage{makeidx}
\usepackage{rwsym}
\usepackage{blkarray}
\usepackage{booktabs}
\usepackage{amsthm}
\usepackage{cite}
\usepackage{psfrag}
\usepackage{color}
\newcommand{\mathstyle}{\scriptstyle}
\begin{document}

\title{Constrained Phase Noise Estimation in OFDM Using Scattered Pilots Without Decision Feedback}
\author{\IEEEauthorblockN{{\Large Pramod Mathecken, Taneli Riihonen, Stefan Werner and Risto Wichman}}
	\thanks{The authors are with Aalto University  School of Electrical Engineering, Department of Signal Processing and Acoustics, P.O.\ Box 13000, FI-00076 Aalto, Finland (Email: \{pramod.mathecken,
		taneli.riihonen,  stefan.werner, risto.wichman\}@aalto.fi).}}
\maketitle
\begin{abstract}
In this paper,  we consider an OFDM radio link corrupted by oscillator  phase noise in the receiver, namely the problem of estimating and compensating for the impairment. To lessen the computational burden and delay incurred onto the receiver, we estimate  phase noise  using only  scattered pilot subcarriers,\ie, no tentative symbol decisions are used in obtaining and improving the phase noise estimate. In particular, the phase noise estimation problem is posed as an unconstrained optimization problem whose minimizer suffers from the so-called \emph{amplitude and phase estimation error}. These errors arise due to receiver noise, estimation from limited scattered pilot subcarriers and estimation using a dimensionality reduction model. It is empirically shown that, at high signal-to-noise-ratios, the phase estimation error is small. To reduce the amplitude estimation error, we restrict the minimizer to be drawn from the so-called \emph{phase noise geometry}  set when minimizing the cost function. The  resulting optimization problem is a non-convex program. However, using the \emph{S-procedure for quadratic equalities}, we show that the optimal solution can be obtained by solving the convex dual problem. We also consider a less complex  heuristic scheme that achieves the same objective of restricting the minimizer to the phase noise geometry  set. Through simulations, we demonstrate improved coded bit-error-rate and phase noise estimation error performance when enforcing the phase noise geometry. For example, at high signal-to-noise-ratios, the probability density function of the phase noise estimation error exhibits \emph{thinner tails}  which results in lower  bit-error-rate.  
\end{abstract}

%
\IEEEpeerreviewmaketitle

\newtheorem{proposition}{Proposition}
\newtheorem{theorem}{Theorem}
\newtheorem{remark}{\emph{Remark}}
\newtheorem{RC}{Regularity condition}
\newtheorem{lemma}{Lemma}
\section{Introduction}

In this paper, we focus on  the phase noise problem in orthogonal frequency division multiplexing (OFDM) which falls in  the category of RF-impairments. It is well known that the OFDM waveform is sensitive to  RF-impairments  which also include power amplifier non-linearities, IQ-imbalance and jitter noise \cite{DirtyRF}. Phase noise refers to random fluctuations in  the phase of the carrier signal that is used for transmission and reception of the baseband information-bearing signal. It arises due to imperfections in the local oscillators that generate the carrier signals. These imperfections exist, simply,  due to the inherent physical nature of these devices but, however, it can be controlled by judicious choice of oscillator design \cite{Rubiola}. 

In the area of performance analysis, plethora  of studies demonstrate a performance drop for an OFDM system corrupted by phase noise \cite{stott, 380034, 5165401, 6188996, 5733452,6472311}. The performance metrics typically used are signal-to-noise-plus-interference-ratio (SINR), bit-error-rate (BER) and channel capacity. The trade-off is typically between the OFDM subcarrier spacing and $3$-dB bandwidth of oscillator power spectral density (PSD) which in turn can be related to the  oscillator topology and circuit parameters \cite{1159110}. A small ratio of  subcarrier spacing and $3$-dB PSD bandwidth results in lower SINR, BER and capacity. These performance studies  were indeed  extended to include other kinds of RF-impairments which are mainly  IQ-imbalance, power amplifier non-linearities and jitter noise \cite{Anttila_phd}.   Numerous algorithms are available that remove phase noise from the received OFDM signal. These algorithms typically require knowledge of the  channel. Some of the state-of-the-art methods on channel estimation in the presence of phase noise  can be found in  \cite{1677918, 5508318,  4567677, 4568453, 6297481, 6868950}. 

The phase noise estimation algorithms can be broadly classified into three types: decision-feedback-based schemes also known as decision-directed  algorithms \cite{7366606,6868950,syryala_crowncomp,5963413, 4568453, 4291833,4027596}; pilot-based  schemes that use the scattered pilot structure provided in  LTE \cite{5508318, 5068965, 1033877}; and, finally, blind estimation schemes \cite{4291878, 4156406}. Decision-feedback schemes estimate phase noise using  tentative decisions on the transmitted symbols. Using the obtained estimate, phase noise is removed and new decisions on the transmitted symbols are taken which are again used to refine the phase noise estimate. The process is iterated over a certain number of times, thus, resulting in a feedback loop. Because of this iteration procedure,  these schemes can impose a significant computational burden onto the receiver. The primary goal in blind estimation schemes is to  jointly estimate phase noise and  transmitted symbols. These approaches typically use Bayesian filtering methods to jointly estimate the desired parameters \cite{Simo}. For example, in  \cite{4156406}, \emph{variational-inference}  is used, while  \emph{Monte-Carlo} methods are used in  \cite{4291878}. These methods, although statistically optimal, are computationally intensive and may not be suitable in delay-sensitive wireless systems.  

Pilot-based schemes that utilize scattered pilot subcarriers provide a computationally attractive alternative to decision-feedback and blind estimation schemes. There exists plethora of work where, using scattered pilot subcarriers, only the \emph{common phase error} (CPE) is estimated while the higher-order frequency components of phase noise, also known by \emph{inter-carrier-interference} (ICI), are assumed to be small and, hence, not estimated \cite{1356210, 1425753, 4567661}. It is well known that, for satisfactory performance, the ICI must also be estimated. To the best of the authors knowledge,  \cite{5508318, 5068965} and \cite{1033877} are the only available works that, using only scattered pilot subcarriers,  estimate both  CPE and ICI terms.  One of the goals of this paper is to contribute towards scattered pilot-based phase noise estimation schemes that estimate both  CPE and ICI terms  with high degree of accuracy.

In this paper, for phase noise estimation, we use two new aspects of phase noise  that have been recently discovered: The first is the so-called \emph{phase noise spectral geometry}; and second is a new dimensionality reduction model that preserves this geometry when moving from lower to higher dimensional spaces. These two aspects of phase noise were originally proposed in \cite{7366606}, however, used in developing a decision-feedback phase noise estimation scheme which has  high computational complexity. We build upon these ideas to develop a novel scattered pilot-based estimation scheme without any decision feedback loop. We show in this work that utilizing the phase noise spectral geometry in conjunction with this new dimensionality reduction model  improves the estimation error performance and, hence, the  BER.

The main contributions of this paper are as follows:
\begin{itemize}
\item As our starting point, we use the least-squares (LS) approach of \cite{1033877} to estimate the desired phase noise spectral vector using scattered pilot subcarriers. We show that the minimizer of the resulting unconstrained optimizaton problem suffers from \emph{amplitude and phase estimation errors} which arises due to receiver noise, estimation from limited scattered pilot subcarriers and estimation using a dimensionality reduction model.  We empirically show that, at high SNRs, the phase estimation error is small and the critical factor is the amplitude estimation error. 
\item To eliminate  the amplitude estimation error, we impose the phase noise geometry as constraints when minimizing our cost function. The resulting optimization problem is a non-convex program, and we show using the so-called \emph{S-procedure} that the optimization problem can be solved equivalently using the convex dual problem. We also present a heuristic scheme with reduced computational complexity that  achieves the same objective of enforcing the estimate to satisfy the phase noise geometry. 
\item We provide conditions for the S-procedure to be lossless for generic quadratic equalities. In \cite{7366606}, the authors present the S-procedure for quadratic equalities specific to their problem. In this paper, we  build upon the ideas presented in \cite{7366606} and generalize the S-procedure for generic  quadratic equalities. We  use the S-procedure to prove optimality of our proposed optimization problem. 
\end{itemize}

The paper is structured as follows: In Section \ref{sec:sysmod}, we present the OFDM system model impaired by phase noise. This shall serve as the foundation for the rest of the paper. Section \ref{sec:Background} covers two particular aspects: The first aspect summarizes the findings of \cite{7366606} which are the phase noise spectral geometry and the phase noise geometry-based dimensionality reduction model. The second aspect dwells on the topic of S-procedure for generic quadratic equalities. We  use the S-procedure in later sections to prove optimality of the proposed phase noise optimization problem. Section \ref{sec:PNschemes} presents the proposed scattered pilot-based phase noise estimation schemes. Specifically, two new schemes are proposed with the first being the optimal scheme while the second scheme is heuristic in nature, however, with reduced computational complexity. In  Section \ref{sec:NR}, we present  numerical results of the proposed estimation schemes.
\section{System Model}
\label{sec:sysmod}
In an  OFDM system, an information symbol vector, denoted by $\bbs = \tr{\left[s_0~s_1\ldots s_{\nc-1} \right]}$,  is transmitted using $\nc$ orthogonal subcarriers \cite{pun}. These subcarriers pass through a frequency-selective channel whose  discrete-time impulse response is denoted by  $h[n]$. At the receiver side, the signal gets corrupted by the receiver additive noise and phase noise. Assuming sufficient timing synchronization, the received symbol vector is given by
\begin{align}
\label{eq:sysmodvec}	
\bbr=\bbV\bbH\bbs + \bbn,
\end{align}
where $\bbH$ is a diagonal matrix composed of elements $\{H_k\}_{k=0}^{N_c-1}$ which are the discrete Fourier transform (DFT) of  $h[n]$,\ie,
\begin{align}
\label{eq:channelk}	
H_k=\sum_{n=0}^{\nc-1}h[n]e^{-\jmath(2\pi k n)/N_c}, k=0,1,\ldots,\nc-1.
\end{align} 
The vector $\bbn$ denotes the additive receiver noise which is Gaussian with diagonal covariance matrix whose diagonal values are equal to $\sigma_{\rm n}^2$.  The effect of phase noise is represented by the unitary matrix $\bbV$ which is row-wise circulant with the first row vector being $\trh{\bbdelta}$ which denotes Hermitian transpose of the column vector $\bbdelta$. The elements of $\bbdelta$ are given by 
\begin{align}
\label{eq:deltak}	
\delta_k=\sum_{n=0}^{\nc-1}\frac{e^{-\jmath\theta[n]}}{N_c}e^{-\jmath(2\pi k n)/N_c}, k=0,1\ldots,\nc-1,
\end{align}
where $\theta[n]$ is the receiver phase noise.  In this paper, we  refer to $\bbdelta$ as the \emph{phase noise spectral vector}. 

Ideally, in the absence of phase noise (\ie, when $\theta[n]=0$) and after using \eqref{eq:deltak}, we have  $\bbdelta = \tr{[1,0,\ldots,0]}$ and, hence, $\bbV = \bbI_{\nc}$, where $\bbI_{\nc}$ denotes the $\nc \times \nc$
identity matrix. Equation \eqref{eq:sysmodvec}, thus,  reduces to $\bbr = \bbH\bbs + \bbn$ which is the standard OFDM system model with no phase noise. In practice, phase noise is always present which renders $\bbV$ to constitute  non-zero off-diagonal elements. 
\section{Background: Phase Noise Spectral Geometry, Dimensionality Reduction and S-Procedure}
\label{sec:Background}
In this section, we dwell on three particular topics which shall be used in later sections to develop   phase noise estimation schemes. In Section \ref{sec:Geometryofdelta}, we  present  the  geometry of $\bbdelta$, while in Section \ref{sec:dimensionalitymodel}, we present a new dimensionality reduction model that takes into account this geometrical aspect of $\bbdelta$. Finally  in Section \ref{sec:Sprocedure}, we present the  S-procedure for quadratic equalities which shall be used to prove optimality of one of our phase noise estimation schemes.  The results in   Sections \ref{sec:Geometryofdelta} and \ref{sec:dimensionalitymodel}  were  originally derived in \cite{7366606} and, hence, we summarize the main points. The S-procedure for quadratic equalities in Section \ref{sec:Sprocedure} is a generalization of the approach used in  \cite{7366606} which was limited to quadratic equations specific to their application. 
\subsection{Geometry of $\bbdelta$}
\label{sec:Geometryofdelta}
From \eqref{eq:deltak}, we see that $\bbdelta_k$ is the DFT of $\frac{e^{-\jmath\theta[n]}}{\nc}$ which has \emph{constant-magnitude} time-domain samples. Intuitively, we could expect this time-domain property to manifest in the frequency domain in some equivalent form. This is indeed the case which is easy to show and derived in \cite{7366606}. Specifically, it is shown that $\bbdelta$ always satisfies
\begin{equation}
\label{eq:sph_base}
\trh{\bbdelta}\bbP_l\bbdelta 	= \Lambda_l,~l=0,1\ldots,\nc-1,
\end{equation}
where $\Lambda_l$ is the Kronecker delta function,\ie, $\Lambda_0 = 1$ and $\Lambda_l=0,~l=1,2,\ldots,\nc-1$. The matrix $\bbP_l = \left(\bbP_1\right)^l$ is a  permutation matrix defined by the $\nc \times \nc$ matrix $\bbP_1$.  The first column of  $\bbP_1$ is given by the $\nc \times 1$ vector $\tr{\left[0,1,0,\ldots,0\right]}$ and the $j$-th column is obtained by circularly shifting the vector $j-1$ times to the bottom.   For $l=0$, we get the unit-norm property, where $\bbP_0=\bbI_{\nc}$. 

Equation \eqref{eq:sysmodvec} provides the relation between $\bbr$ and $\bbs$ for any OFDM symbol. For different OFDM symbols, we obtain different realizations of the channel matrix $\bbH$, $\bbV$ and $\bbn$. Thus, although $\bbV$ or $\bbdelta$ vary from one OFDM symbol to another, from \eqref{eq:sph_base}, we see that $\bbdelta$ is always drawn from a particular set. This is useful from an estimation point of view because we now know  \emph{where} to look for $\bbdelta$. 
\subsection{Dimensionality Reduction}
\label{sec:dimensionalitymodel}
The effect of phase noise can be \emph{compensated} straightforwardly if we had knowledge of  $\bbdelta$. We can then form the matrix $\bbV$ and perform $\trh{\bbV}\bbr = \bbH\bbs + \trh{\bbV}\bbn$ to remove  phase noise (we use the fact that $\trh{\bbV}\bbV = \bbI_{\nc}$). Thus, the critical task of \emph{estimation} is to obtain this knowledge as accurately as possible using which phase noise can be compensated.
\subsubsection{The Conventional Model}
 From the point of view of estimation, estimating the entire vector $\bbdelta$ may not be feasible since the dimensionality of $\bbdelta$, equal to  $\nc$,  can be large. For example, in LTE, $\nc > 100$, and it can be as large as $2048$. In practice, system specifications enforce stringent requirements on  oscillator performance which effectively result in  tolerable and slow-varying phase noise processes. This has the effect of larger concentration of power in the low frequency components represented by the top and bottom components of $\bbdelta$, while the  high frequency terms represented by the middle components of $\bbdelta$ constitute only a small fraction of total power. We can, thus, \emph{model} $\bbdelta$ as follows:
\begin{align}
\label{eq:selection_mat}
\bbdelta &= 
\begin{pmatrix}
\bbI_{m\times m} &  \bzero_{k\times k } \\
\bzero_{\nc-(m+k) \times m} & \bzero_{\nc-(m+k) \times k}\\
\bzero_{k \times m} &   \bbI_{k\times k}
\end{pmatrix}\bgamma =  \bbL\bgamma,
\end{align}
where $\bzero$ is the matrix of zeros of appropriate dimensions. The matrix $\bbL$ is of dimension $\nc \times N$, $N= m+k$, and $\bgamma$ comprises of the $N$ low-frequency components. Thus, rather than estimating $\bbdelta$, we estimate the smaller $N$-dimensional vector $\bgamma$ and then use \eqref{eq:selection_mat} to finally obtain our estimate of $\bbdelta$. Note that from \eqref{eq:selection_mat}, we set the high-frequency components  to zero. The model in \eqref{eq:selection_mat} is  commonly used in the literature related to phase noise estimation. We shall also refer to $\bbL$ as low frequency transformation matrix or LFT. It is useful and practical especially when the phase noise process is slow-varying. 
Unfortunately, the model of \eqref{eq:selection_mat} does not guarantee that  $\bbdelta$ obtained from   \eqref{eq:selection_mat} will satisfy \eqref{eq:sph_base}. 
\subsubsection{The  Geometry-preserving Model} 
\label{sec:geomodel}
In \cite{7366606}, a new model relating $\bbdelta$ and $\bgamma$ is proposed. This  is given as follows: The vector  $\bbdelta$ acquires its \emph{properties} from a   smaller dimensional phase noise spectral vector $\bgamma$ that satisfies the $N$-dimensional equivalent of \eqref{eq:sph_base},\ie,
\begin{align}
\label{eq:sph_base_lowdim}
\trh{\bgamma}\bbPt_l\bgamma 	= \tilde{\Lambda}_l,~l=0,1,\ldots,N-1, 
\end{align}
where  $\bbPt_l$ and $\tilde{\Lambda}_l$ are the $N$-dimensional equivalents  of $\bbP_l$  and $\Lambda_l$, respectively. The vectors $\bbdelta$ and $\bgamma$ are linearly related as 
\begin{align}
\label{eq:ppt}
\bbdelta &= \bbT\bgamma,
\end{align}
where the $\nc \times N$ matrix $\bbT$  is of the form
\begin{align}
\label{eq:Tmat}
\bbT = \bbF\bbTt\trh{\bbFt},
\end{align}
where the respective $\bbFt$ and $\bbF$ are the $N\times N$ and $\nc \times \nc$ DFT matrices and the columns  $\bbtt_i$ of the $\nc \times N$ matrix $\bbTt$ must satisfy,  for all $l=1,2,\ldots,\nc-1$,
\begin{align}
\label{eq:coll_cond}
\trh{\bbTt}\bbTt = \bbIt,~\trh{\bbtt_i}\bbD_l\bbtt_j = 0~{\rm for~}i \neq j,~\sum_{i=0}^{N-1}\trh{\bbtt_i}\bbD_l\bbtt_i = 0, 
\end{align}
where the diagonal $\bbD_l = \trh{\bbF}\bbP_l\bbF$.  In comparison with the conventional model of \eqref{eq:selection_mat}, the geometrical model imposes restrictions on $\bgamma$ and the transformation matrix $\bbT$. The role of $\bbT$ is to preserve the phase noise geometry when moving from lower to higher dimensional spaces.  Because of the geometry preserving nature of $\bbT$, we shall refer to it as the \emph{phase noise geometry preserving transformation} or PPT. In reality, many possible choices of  PPT exists and in the following paragraph, we provide  one such  example that we shall later use. 

\paragraph{Piecewise constant PPT (PC-PPT)}
The transformation $\bbdelta = \bbF\bbTt\trh{\bbFt}\bgamma$ can be interpreted as follows: $\trh{\bbFt}\bgamma$ is a $N$-dimensional time-domain vector which is interpolated (by $\bbTt$) to a higher dimensional vector and then transformed to the Fourier domain. Such an interpretation is valid for  phase noise since, in general, it is a low-pass process. One of the simplest interpolators is to simply repeat the elements of the time-domain vector,\ie, 
\begin{align}
\label{eq:PCPPT}
\bbTt_{\rm pc} = \sqrt{\frac{\nc}{N}}
\begin{pmatrix}
\bone_{\frac{\nc}{N}} & \bzero & \ldots  & \bzero \\
\bzero & \bone_{\frac{\nc}{N}} & \ddots & \vdots \\
\vdots &  \ddots& \ddots & \vdots\\
\bzero &  \ldots & \bzero & \bone_{\frac{\nc}{N}}
\end{pmatrix},
\end{align} 
where $\bone_{\frac{\nc}{N}}$ is an  $\frac{\nc}{N} \times 1$   vector of ones and $\bzero$ is the vector with elements equal to zero. We assume without loss of generality that $\frac{\nc}{N}$ is even. It can be easily verified that $\bbTt_{\rm pc}$  satisfies the conditions of \eqref{eq:coll_cond} and, hence, $\bbT_{\rm pc} = \bbF\bbTt_{\rm pc }\trh{\bbFt}$ is a PPT. 
\subsection{S-procedure for Quadratic Equalities}
\label{sec:Sprocedure}
The S-procedure is a method of replacing a set of quadratic inequalities or equalities  with a  \emph{linear matrix inequality} (LMI). It is typically used when solving  primal and dual optimization problems \cite{Boyd_CO}. In this paper, we concern ourselves with only quadratic equalities. A good overview of the topic for quadratic inequalities can be found in \cite{Jonnson_SP}. 

Consider the following quadratic forms:
\begin{align}
\label{eq:quadraticform}
q_l(\bbx) = \trh{\bbx}\begin{pmatrix}
\bbA_l & \bbd_l \\
\trh{\bbd_l} & c_l
\end{pmatrix}\bbx,l=0,1,\ldots,L-1,
\end{align}
where $\bbx \in \calC^{N+1}$. Define the sets:
\begin{align}
\label{eq:setQ}
\cQ &= \Big\{\tr{\Big(q_0(\bbx), q_1(\bbx),\ldots,q_{L-1}(\bbx)\Big)}:  \bbx \in \calC^{N+1} \Big\},  \\
\cN &= \left\{ \tr{\left(g,\tr{\bzero}_{L-1}\right)}~{\rm s.t}~g < 0\right \}, \label{eq:setN}
\end{align}
where $\bzero_{L-1}$ is a $L-1\times 1$ vector of zeros.
Now consider the following two statements:
\begin{itemize}
	\item S1: $q_0(\bbx) \geq 0$ whenever $ q_l(\bbx) = 0$ for all $l > 0$. This is equivalent to $\cQ \cap \cN = \emptyset$, where $\cap$ denotes intersection and $\emptyset$ denotes the empty set. 
	\item S2: There exists constants $\rho_l,l=1,2,\ldots,L-1$ such that 
	\begin{align}
	\label{eq:LMI}
\bbAt =	\begin{blockarray}{cc}
	\begin{block}{(cc)}
	\bbA_0 + \sum_{l=1}^{L-1}\rho_l\bbA_l   &  \bbd_0 +\sum_{l=1}^{L-1}\rho_l\bbd_l \\
\trh{(\bbd_0 +\sum_{l=1}^{L-1}\rho_l\bbd_l)}	& c_0+\sum_{l=1}^{L-1}\rho_lc_l  \\
	\end{block} 
	\end{blockarray} \succeq 0.
	\end{align}
\end{itemize}
We say that the S-procedure is lossless if the statements S1 and S2 are equivalent,\ie, S1 implies S2 and S2 implies S1. We now have the following Lemma:
\begin{lemma}
	\label{lemma:S2toS1}
S2 always implies S1.	
\end{lemma}
\begin{IEEEproof}
S2 implies that, for all $\bbx \in \calC^{N+1}$,  $\trh{\bbx}\bbAt\bbx \geq 0$ and after using  the expression of $\bbAt$,  
\begin{align}
&q_0(\bbx) +\sum_{l=1}^{L-1}\rho_lq_l(\bbx) \geq 0  \\
&\tr{\brho}\bby \geq 0, {\rm for}~\bby \in \cQ \label{eq:rhsconQ},
\end{align}
where $\brho =\tr{\left[1,\rho_1,\rho_2,\ldots,\rho_{L-1} \right]}$. For such a  $\brho$, we also have
\begin{align}
\tr{\brho}\bby = g < 0, {\rm for}~\bby \in \cN  \label{eq:conN},
\end{align}
which results from the definition of $\cN$. Thus, from \eqref{eq:conN} and \eqref{eq:rhsconQ}, we see that $\cQ \cap \cN = \emptyset$ which is equivalent to  S1.
	\end{IEEEproof}

Unfortunately, S1 does not necessarily imply S2, and only depending upon the type of the set $\cQ$ it may imply S2. By imposing a certain type of structure on $\cQ$, the implication of S1 to S2 can be achieved. The following \emph{regularity condition}  imposes such a structure on $\cQ$. First, define the set 
\begin{align}
\cQt = \Big\{\bbq(\bbx)=\tr{\Big(q_1(\bbx),  q_2(\bbx),\ldots,q_{L-1}(\bbx)\Big)}: \bbx \in \calC^{N+1} \Big\}.
\end{align}
We form a matrix 
\begin{align}
\label{eq:Qmat}
\bbQ =\left[\bbq(\bbx_1)~ \bbq(\bbx_2)~\bbq(\bbx_3) \ldots \bbq(\bbx_M)\right],
\end{align}
for some $\{\bbx_i\}_{i=1}^{M}$. 
\begin{RC}
	There exists vectors $\{\bbx_i\}_{i=1}^{M} \neq \bzero $, where $M > L-1$, and constants $\{p_i\}_{i=1}^{M} > 0$ such that 
	\begin{align}
	&{\rm \bf rank}\Big{(}\bbQ\Big{)} = L-1, \label{eq:R2} \\
	&\sum_{i=1}^{M}p_i\bbq(\bbx_i) = \bzero. ~\label{eq:R1}
	\end{align}
\end{RC}
\begin{remark}
	\label{rem:RC}
	The regularity condition implies that there does not exist any hyperplane passing through the origin such that all points $\{\bbq(\bbx_i)\}_{i=1}^{M}$ lie on one side of the hyperplane.  This is seen as follows: For any non-zero $\bbat \in \calR^{L-1}$, taking the inner product\wrt\ $\bbat$ on both sides of \eqref{eq:R1}, we have $\sum_{i=1}^{M}p_i(\tr{\bbat}\bbq(\bbx_i)) = 0 $  which implies that $\tr{\bbat}\bbq(\bbx_i) \geq 0 $ or $\tr{\bbat}\bbq(\bbx_i) \leq 0$ for all $i=1,2,\ldots, M$ is not possible since  $\{p_i\}_{i=1}^{M} > 0$. The special case of $\tr{\bbat}\bbq(\bbx_i) = 0$ for all $i=1,2,\ldots, M$ implies ${\rm \bf rank}\left(\bbQ\right) < L-1$ which contradicts with  \eqref{eq:R2}.  Hence, for any non-zero $\bbat$, we must have 
	\begin{align}
	\label{eq:remark1}
	\tr{\bbat}\bbq(\bbx_i) < 0,~\tr{\bbat}\bbq(\bbx_j) > 0~ 
	{\rm for~some~}i {\rm~and~}j, i\neq j. 
	\end{align} 
\end{remark}
\begin{remark}
The regularity 	condition also implies that the conic hull of $\cQt$  is equal to $\cR^{L-1}$. This follows from Remark \ref{rem:RC}.
\end{remark}
We now have the following theorem on the losslessness of the S-procedure.
\begin{theorem}
	\label{appen:theorem}
	Assume $\cQt$ satisfies the regularity condition. Let  ${\rm cov}\left(\cQ\right)$ denote the convex hull of $\cQ$. If $\cQ \cap \cN =  \emptyset$ implies ${\rm cov}\left(\cQ\right) \cap \cN =  \emptyset$ then the S-procedure is lossless.
\end{theorem}
\begin{IEEEproof}
First, we note that $\cQ \cap \cN =  \emptyset$ implies the sets are disjoint. Also, the sets $\cN$ are ${\rm cov}\left(\cQ\right)$ are convex sets. Thus, if $\cQ \cap \cN =  \emptyset$ implies ${\rm cov}\left(\cQ\right) \cap \cN =  \emptyset$   then there exists a hyperplane \emph{passing through the origin} that separates ${\rm cov}\left(\cQ\right)$  and $\cN$ \cite{Boyd_CO,Nemirovski_CO},\ie, there exists constants $a_l$ such that
\begin{align}
\bba^T\bby & \leq 0,~\bby \in \cN,  \label{eq:sephyp_side1} \\
\bba^T\bby &\geq 0,~\bby \in {\rm cov}(\cQ), \label{eq:sephyp_side2}
\end{align}
where $\bba = \tr{\left[a_0, a_1,\ldots,a_{L-1}\right]}$. From \eqref{eq:sephyp_side1} and  definition of $\cN$, we must have $a_0 \geq 0$. Now $a_0=0$ is impossible because of the regularity condition assumption. This is seen as follows: First, define the vector $\bbat$ with components as $\{a_i\}_{i=1}^{L-1}$. Assume $a_0 = 0$ is true. Then at points $\tr{[q_0(\bbx_i)~\tr{\bbq(\bbx_i)}]} \in {\rm cov}(\cQ)$ with $\{\bbx_i\}_{i=1}^{M}$  as defined in the regularity condition, \eqref{eq:sephyp_side2} becomes
\begin{align}
\label{eq:contradict}
\tr{\bbat}\bbq(\bbx_i) \ge 0, {\rm for~all~} i=1,2,\ldots,M.
\end{align}
Equation \eqref{eq:contradict} contradicts with \eqref{eq:remark1} of Remark \ref{rem:RC} which is satisfied because of the regularity condition assumption.  Hence, $a_0 > 0$ is necessary. Hence, ${\rm{for~all~}}\bbx \in \calC^{N+1}$, \eqref{eq:sephyp_side2} implies 
\begin{align}
q_0(\bbx) + \sum_{l=1}^{L-1}\frac{a_l}{a_0}q_1(\bbx) \geq 0.
\end{align}   
Writing $\rho_l = \frac{a_l}{a_0}$, and  after substituting the expressions of $q_l(\bbx)$ we obtain \eqref{eq:LMI},\ie, S1 implies S2. After using Lemma \ref{lemma:S2toS1}, we have S1  equivalent to S2.
	\end{IEEEproof}
\section{Phase Noise Estimation Schemes}
\label{sec:PNschemes}
In this section, we present  scattered pilot-based phase noise estimation schemes  that take into account the phase noise spectral geometry.  In \cite{1033877}, the authors estimate  $e^{-\jmath \theta[n]}$ from scattered pilots using the LS approach. We can equivalently apply the  same approach in the frequency domain for estimation of $\bbdelta$. Through  error analysis,  we show that the derived  LS estimator suffers from amplitude and phase  estimation errors.  We  improve the scheme by  enforcing the  phase noise geometry as constraints when minimizing the LS cost function. 
\subsection{Unconstrained LS (ULS) Estimation of \cite{1033877} }
Denote that  $\bbw = \bbH\bbs$. We assume  knowledge of the diagonal channel matrix $\bbH$. 
Let $\bbw_{\rm p}$ denote the $ K \times 1$ vector of pilot subcarrier symbols which can be obtained from $\bbw$ as 
\begin{align}
\bbwp = \bbK\bbw,
\end{align}
where the rows of the  $K \times \nc$ matrix  $\bbK$ are orthogonal and given by the unit-vectors $\tr{\bbe_j} = \left[0,\ldots,0,1,0,\ldots,0\right], j \in \set{1,2,\ldots,\nc}$. Let $\hat{\bbV}$ denote our estimate of the matrix $\bbV$. An estimate of $\bbw_{\rm p}$ can be obtained from \eqref{eq:sysmodvec} as
\begin{align}
\bhwp = \bbK\trh{\hat{\bbV}}\bbr = \bbK\bbR\bhdelta, 
\end{align}
where $\bbR$ is column-wise circulant with the first column vector $\bbr$. The $j$-th column of $\bbR$ is obtained by circularly shifting $\bbr$ $j-1$ times to the bottom. It results from the assumption that   $\trh{\hat{\bbV}}$ is unitary circulant with the column vector $\bhdelta$. We use a basis set $\bbB$ to represent $\bbdelta$,\ie,
\begin{align}
\label{eq:basisrepdel}
\bbdelta  = \bbB\balpha = \bbT\bgamma + \bbU\bbeta.
\end{align}
Let $\bhgamma$ denote our estimate  of $\bgamma$. Then our estimate of $\bbdelta$ is  
\begin{align}
\label{eq:delestimate}
\bhdelta = \bbT\bhgamma.
\end{align}
Essentially, the term $\bbU\bbeta$ in \eqref{eq:basisrepdel} represents the unestimated part of $\bbdelta$. A good choice of $\bbB$ is when most of the power is in the $\bgamma$ term.   The estimate   $\bhgamma$ can now be obtained from $\bbwp$ by minimizing the LS error between $\bbwp$ and $\bhwp$,\ie, 
\begin{align}
\label{eq:lscost}
\calJ(\bhgamma) &= \norm{\bbK\bbR\bbT\bhgamma - \bbwp}_2^2 \\
&= \trh{\bhgamma}\bbM\bhgamma -\trh{\bhgamma}\bbb-\trh{\bbb}\bhgamma + \trh{\bbb}\bbb,
\end{align}
where $\bbM = \trh{\bbT}\trh{\bbR}\trh{\bbK}\bbK\bbR\bbT$ and $\bbb = \trh{\bbT}\trh{\bbR}\trh{\bbK}\bbwp$. 
The minimizer to the above cost function is given by
\begin{align}
\label{eq:gamest}
\bhgamma = \inv{\bbM}\bbb,
\end{align}
and, after using \eqref{eq:delestimate}, the LS estimate of $\bbdelta$ is given by
\begin{align}
	\label{eq:del_lsestimate}
\bhdelta_{\rm ls} = \bbT\inv{\bbM}\bbb.
\end{align}
\subsubsection{Error Analysis}
\label{sec:error_ana}
In this subsection, we shall see the how the LS estimate of \eqref{eq:del_lsestimate} is affected  by: dimensionality reduction represented by $\bbT$;  limited  scattered-pilot knowledge represented by  $\bbK$; and by receiver noise which is embedded in  $\bbR$. The overall effect is introduction  of   amplitude and phase estimation errors in the LS estimate. 

First, we observe that the circulant matrix $\bbR$ is given by 
\begin{align}
\bbR  &= \bbF\diag\left(\trh{\bbF}\bbr\right)\trh{\bbF} \label{eq:Rcirmata} \\
	  &=  \bbF\diag\left(\bbE_{\theta}\trh{\bbF}\bbw + \trh{\bbF}\bbn\right)\trh{\bbF} \label{eq:Rcirmatb} \\	
  	  &=  \bbF\left(\bbE_{\theta}\bbE_{\rm w} +  \bbE_{\rm n}\right)\trh{\bbF} \label{eq:Rcirmatc}	 \\
  	  &=  \bbF\bbE_{\theta}\bbE_{\rm w}\left(\bbI_{\nc} +  \inv{\bbE_{\theta}}\inv{\bbE_{\rm w}}\bbE_{\rm n}\right)\trh{\bbF} \label{eq:Rcirmatd} \\
    &=  \bbF\bbE_{\theta}\bbE_{\rm w}\bbE_{\rm snr}\trh{\bbF}, \label{eq:Rcirmate} 
\end{align}
where $\diag\left(\bbx\right)$ is a diagonal matrix with elements of the vector $\bbx$ as diagonal values.   In \eqref{eq:Rcirmata}, we substitute \eqref{eq:sysmodvec}  and use $ \bbV = \bbF\bbE_{\theta}\trh{\bbF}$ to arrive at \eqref{eq:Rcirmatb}. The diagonal values of the diagonal matrix $\bbE_{\theta}$ are $e^{\jmath\theta[i]}, i=1,\ldots,\nc-1$. We denote as $ \bbE_{\rm w}  = \diag\left(\trh{\bbF}\bbw \right)$, $\bbE_{\rm n} = \diag\left(\trh{\bbF}\bbn \right)$ and $\bbE_{\rm snr} = \bbI_{\nc} +  \inv{\bbE_{\theta}}\inv{\bbE_{\rm w}}\bbE_{\rm n}$ which captures in some sense the SNR. Using \eqref{eq:Rcirmate}	in the expressions for $\bbM$ and $\bbb$ while making use of the representation of $\bbT$ in  \eqref{eq:Tmat}, we can re-write  \eqref{eq:del_lsestimate} as
\begin{align}
\label{eq:lsest_amp_noise}
\bhdelta_{\rm ls}  = \bbF\bbTt\inv{\left(\trh{\bbTt}\trh{\bbE_{\rm w}}\trh{\bbE_{\rm snr}}\bbP_{\rm r}\bbE_{\rm snr}\bbE_{\rm w}\bbTt \right)}\trh{\bbTt}\trh{\bbE_{\rm w}}\trh{\bbE_{\rm snr}}\trh{\bbE_{\theta}}\trh{\bbF}\trh{\bbK}\bbw_{\rm p},
\end{align}
where the projection matrix $\bbP_{\rm r} = \trh{\bbE_{\rm \theta}}\trh{\bbF}\trh{\bbK}\bbK\bbF\bbE_{\theta}$.  
Writing as $\bbE_{\rm p} = \diag(\trh{\bbF}\trh{\bbK}\bbw_{\rm p})$ in \eqref{eq:lsest_amp_noise} and using the fact that the diagonal values of $\trh{\bbE_{\rm \theta}}$ take the form $e^{-\jmath \theta[i]}$, we finally obtain
\begin{align}
\label{eq:lsest_amp_noisef}
\bhdelta_{\rm ls}  = \bbF\bbC\trh{\bbF}\bbdelta,
\end{align}
where the $\nc \times \nc$  matrix $\bbC$ is given by
\begin{align}
\label{eq:Amat}
\bbC = \bbTt\inv{\left(\trh{\bbTt}\trh{\bbE_{\rm w}}\trh{\bbE_{\rm snr}}\bbP_{\rm r}\bbE_{\rm snr}\bbE_{\rm w}\bbTt \right)}\trh{\bbTt}\trh{\bbE_{\rm w}}\trh{\bbE_{\rm snr}}\bbE_{\rm p}. 
\end{align}

In the ideal case, we would like $\bbC = \bbI_{\nc}$ which would render complete  knowledge of $\bbdelta$. However, the following reasons prevent $\bbC$ from being the identity matrix:
\begin{itemize}
\item Effect of dimensionality reduction: When $N < \nc$  we have, in general, $\rank\left(\bbT\right) = \rank\left(\bbTt\right) = N$. Thus,  when $N \leq K $ and for any  choice of $\bbK$, $\bbE_{\rm w}$ and $\bbE_{\rm snr}$, we have that  $\rank\left(\bbC\right) = N$. 
\item Effect of receiver noise: This is captured by $\bbE_{\rm snr}$. For example, in the case  when $N=\nc$ and $\bbK = \bbI_{\nc}$, we have $\bbP_{\rm r}= \bbI_{\nc}$, $\bbE_{\rm p} = \bbE_{\rm w}$ and \eqref{eq:lsest_amp_noisef} reduces to
\begin{align}
\bhdelta_{\rm ls} = \bbF \inv{\bbE_{\rm snr}}\trh{\bbF}\bbdelta.
\end{align}
From the expression of $\bbE_{\rm snr}$, we observe that in the presence of receiver noise, in general,   $\inv{\bbE_{\rm snr}} \neq \bbI_{\nc}$. 
\item Effect of scattered-pilots: The quantity $K$ denotes the number of scattered-pilot subcarriers.  The LS estimation of the $N \times 1$ vector $\bhgamma$  using $K$ scattered-pilot subcarriers imposes the inequality $N \leq K < \nc$. This results in $\rank\left(\bbC\right) = N$. 
\end{itemize}
The non-identity nature of $\bbC$ introduces  amplitude and phase estimation errors which is seen as follows: Let $c_{ij}$ denote the $(i,j)$th element of  $\bbC$ and $\hat{\bbx}_{\rm ls} = \trh{\bbF}\bhdelta_{\rm ls}$. We then have 
\begin{align}
\hat{\bbx}_{\rm ls}[i] &= \frac{e^{-\jmath \theta[i]}}{\nc}\left(\sum_{j=0}^{\nc-1}c_{ij}e^{\jmath(\theta[i]-\theta[j])}\right) 	\label{eq:ampnoisea} \\
						&=  \frac{\kappa[i]}{\nc}e^{-\jmath\left(\theta[i] - \omega[i]\right)}, 	\label{eq:ampnoiseb}
\end{align}
where  $\kappa[i] = \abs{\left(\sum_{j=0}^{\nc-1}c_{ij}e^{\jmath(\theta[i]-\theta[j])}\right)}$  and $\omega[i] = \arg{\left(\sum_{j=0}^{\nc-1}c_{ij}e^{\jmath(\theta[i]-\theta[j])}\right)}$ is the phase estimation error.   The amplitude estimation error is given by $\varepsilon[i] = 1 - \kappa[i]$ since ideally $\kappa[i] = 1$. The total estimation error is given by
\begin{align}
\label{eq:estimation_error}
\sum_{i=0}^{\nc-1}\abs{\hat{\bbx}_{\rm ls}[i] - \frac{e^{-\jmath \theta[i]}}{\nc}}^2 &=  \frac{1}{\nc^2}\Big{[}\sum_{i=0}^{\nc-1}(\varepsilon[i])^2 + 2\varepsilon[i](1-\cos(\omega[i])) \nonumber \\
&+ 2\left(\nc - \sum_{i=0}^{\nc-1}\cos(\omega[i]) \right)\Big{]}.
\end{align}

\begin{figure}[!t]
	\begin{center}
		\psfrag{Prob}[c]{{\footnotesize Probability density}}
		\psfrag{real}[c]{{\footnotesize $\omega$ }}
		\includegraphics[height=0.3\textheight, width=0.47\textwidth]{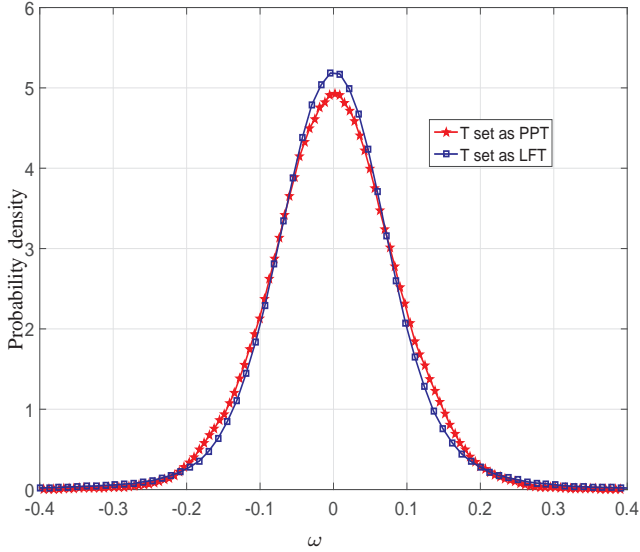}
		\caption{Empirical PDF of the phase estimation error $\omega$ at SNR equal to $30$-dB. The respective PPT and LFT matrix used are  $\bbT_{\rm pc}$ of \eqref{eq:PCPPT} and $\bbL$ of \eqref{eq:selection_mat}.}
		\label{fig:Omegapdf_snr_30}
	\end{center}
\end{figure}

From \eqref{eq:estimation_error}, we see that the estimation error is more sensitive to $\varepsilon[i]$ than $\omega[i]$. This is because it varies quadratically with  $\varepsilon[i]$ and, hence, can grow unbounded, while the variation with  $\omega[i]$ is bounded because of the limited range of the cosine function. The estimation error is minimum at the values $\varepsilon[i] = 0 $ (implies $\kappa[i] = 1$) and $\omega[i] = 0$.  Thus, assuming $\omega[i]$ to be small,  one way of improving the estimation error is to  ensure $\kappa[i] = 1$ which results in $\varepsilon[i] = 0$. For example, we can normalize the samples of $\hat{\bbx}_{\rm ls}[i]$ which ensures that $\kappa[i] = 1$. However, this is not the only approach and in the next section, we present an optimal way of ensuring $\kappa[i] = 1$. This approach of improving the estimation error works well only when   $\omega[i]$ is small. We show empirically that, at high SNRs, this is indeed the case.  Figure \ref{fig:Omegapdf_snr_30} shows the empirical probability density funtion (PDF) of $\omega$ at SNR of $30$-dB.  We see that for any choice of $\bbT$, the PDF is highly concentrated  around the value of zero. For example, even at the  low probability value of $\omega = 0.2$, the estimation error in percentage, after setting $\kappa[i] = 1$ in \eqref{eq:estimation_error}, is close to $4\%$. 
\subsection{Geometry-Constrained LS (GLS)  Estimation}
In this section, we present an estimation scheme that eliminates the amplitude estimation error introduced by the matrix $\bbC$.  
To do so, we  utilize the geometrical model of Section \ref{sec:geomodel}. We first require that we choose $\bbT$ to be a PPT. We then enforce   \eqref{eq:sph_base_lowdim} as constraints when minimizing $\calJ(\bhgamma)$. After obtaining an optimal estimate of $\bgamma$,  our estimate of  $\bbdelta$,\ie, $\bhdelta = \bbT\bhgamma$  also satisfies \eqref{eq:sph_base} (since $\bbT$ is a PPT), thereby eliminating the amplitude estimation error. The optimization problem in terms of $\bhgamma$ is given by
\begin{align}
\label{eq:opt_eq}
~~~~~~~~\left(\cP\right):~~&{\rm Minimize}~\calJ(\bhgamma)  \nonumber \\
&{\rm s.t}~~\trh{\bhgamma}\bhgamma = 1,~\trh{\bhgamma}\bbPt_l^{\rm R}\bhgamma 	= 0,~\trh{\bhgamma}\bbPt_l^{\rm I}\bhgamma 	= 0,\nonumber \\ &~~~~~~~~~~~~~~~~~~l=1,2,\ldots,\frac{N-1}{2},
\end{align}
where $\bbPt_l^{\rm R}$ and $\bbPt_l^{\rm I}$ are the real and imaginary parts of $\bbPt_l$ and are given by
\begin{align}
\label{eq:perm_real_imag}
\bbPt_l^{\rm R} = \frac{\bbPt_l+\trh{\bbPt_l}}{2},   \bbPt_l^{\rm I}=\frac{\jmath(\trh{\bbPt_l}-\bbPt_l)}{2}.
 \end{align}
In \eqref{eq:opt_eq}, we have imposed \eqref{eq:sph_base_lowdim} as constraints, however, elaborated the equations in terms of its real and imaginary parts. This is done so because $\trh{\bhgamma}\bbPt_l\bhgamma, l>0$ is a complex function since the  eigenvalues of  $\bbPt_l$ are complex valued. Thus, the constraint $\trh{\bhgamma}\bbPt_l\bhgamma = 0$ can equivalently be expressed in terms of the real and imaginary parts of the quadratic form as done in \eqref{eq:opt_eq}. We also point to the reader that  only half the number of constraints are enforced in \eqref{eq:opt_eq}. This is because the constraint
\begin{align}
\trh{\bhgamma}\bbPt_{l}\bhgamma  = 0~{\rm implies~} \trh{\left(\trh{\bhgamma}\bbPt_{l}\bhgamma\right)} = 0~{\rm implies~}
\trh{\bhgamma}\bbPt_{N-l}\bhgamma = 0,
\end{align}
where we used the fact that $\trh{\bbPt_l} = \bbPt_{N-l}$. The implication also works in the opposite direction. In \eqref{eq:opt_eq}, we assume that $N$ is odd without any loss in generality.

The optimization problem $\left(\cP\right)$  is typically referred to as the \emph{primal problem}. From \eqref{eq:opt_eq}, we observe  that the constraints are \emph{non-convex} in nature. For example, the unit-norm constraint $\trh{\bhgamma}\bhgamma = 1$ describes, mathematically, an $N$-dimensional sphere, and such an object is a non-convex set. The remaining constraints are also non-convex because the matrices in \eqref{eq:perm_real_imag} constitute both positive and negative eigenvalues. The eigenvalues of $\bbPt_l$ are  $\{e^{\jmath\frac{2\pi nl}{N}}\}_{n=0}^{N-1}$ and, hence, the eigenvalues of $\bbPt^{\rm R}_l$ and $\bbPt^{\rm I}_l$ are $\{\cos(\frac{2\pi nl}{N})\}_{n=0}^{N-1}$ and $\{\sin(\frac{2\pi nl}{N})\}_{n=0}^{N-1}$, respectively. This non-convexity of the constraints renders $\left(\cP\right)$  to be a \emph{non-convex program}.  Most algorithms used in solving non-convex programs yield  local optimal solutions. 
\subsubsection{The Convex Dual Problem} 
A \emph{suboptimal} solution can be obtained by solving the so-called  \emph{dual problem} to $\left(\cP\right)$. It  can be easily derived and is given by \cite{Boyd_CO}
\begin{align}
\label{eq:dual_eq}
\left(\cD \right):~~&{\rm Maximize}~~ \tau \nonumber \\
& {\rm s.t}~\begin{pmatrix}
 \bbM + \lambda\bbI_{N} + \sum_{l=1}^{\frac{N-1}{2}} \alpha_l\bbPt_l^{\rm R} + \beta_l\bbPt_l^{\rm I} & \bbb \\
 \trh{\bbb} & -\tau -\lambda
\end{pmatrix}  \succeq 0,
\end{align}
where $\tau, \lambda, \alpha_l$ and $\beta_l$ are the variables to optimize. In general, the dual problem yields an optimal value different from that of the primal problem (in fact, it is never greater). The dual problem is always a \emph{convex program} which have the property that every local optimal solution is also a global solution. This property eases the search process for algorithms and, in fact, numerous and efficient algorithms exist that solve convex programs in  polynomial time. In certain situations, the dual problem can yield the same optimal value as the primal problem,\ie, a difficult non-convex program can be equivalently solved using an easier convex dual program.

Let $\tau^\diamond, \lambda^\diamond, \alpha^\diamond_l$ and $\beta^{\diamond}_l$ be the minimizer to $\left(\cD \right)$. We obtain our suboptimal estimate of $\bgamma$ by solving the Karhush-Kuhn-Tucker (KKT) necessary condition for local optimality of $\left(\cP\right)$ which is given by
\begin{align}
&\left(\bbM + \lambda^\diamond\bbI_{N}  + \sum_{l=1}^{\frac{N-1}{2}}\alpha^\diamond_l\bbPt^{\rm R}_l + \beta^{\diamond}_l\bbPt^{\rm I}_l\right)\bhgamma_{\rm gls} = \bbb  \label{eq:gammads} \\
&{\rm implies}~\bhgamma_{\rm gls} = \pinv{\left(\bbM + \lambda^\diamond\bbI_{N} + \sum_{l=1}^{\frac{N-1}{2}}\alpha^\diamond_l\bbPt^{\rm R}_l + \beta^{\diamond}_l\bbPt^{\rm I}_l\right)}\bbb  \label{eq:gammads2}
\end{align}
where $\pinv{\bbX}$ denotes pseudo-inverse of $\bbX$. The minimizers  $\tau^\diamond, \lambda^\diamond, \alpha^\diamond_l$ and $\beta^{\diamond}_l$ are obtained by solving  $\left(\cD \right)$ which is a \emph{semi-definite} program (SDP) \cite{Boyd_CO}. SDPs are convex programs and  efficiently solved using interior point algorithms \cite{Nemirovski_IP}. Standard solvers are  available that solve for such programs, for example, in this paper, we use CVX, a package for solving convex programs \cite{Grant2008, Grantcvx}.

Denote the respective optimal values of $\left(\cP\right)$ and $\left(\cD\right)$ by $p^\star$ and $d^\star$. We say the dual problem yields a  \emph{suboptimal} solution whenever $d^\star \leq p^\star$. Such a situation is  referred by the term \emph{weak duality}. When  $d^\star = p^\star$, also known by the term \emph{strong duality}, the optimal solution is equivalently achieved by solving the dual problem. In the next paragraph, we dwell on  when $d^\star = p^\star$ and  show that strong duality holds for the optimization problems $\left(\cP\right)$ and $\left(\cD\right)$. 
\subsubsection{Strong Duality Between $\left(\cP\right)$ and $\left(\cD \right)$}
In this section, we shall use the S-procedure described in Section \ref{sec:Sprocedure} for  proving strong duality between the primal and dual problems. For our application, we set the matrices in \eqref{eq:quadraticform} as follows:
\begin{align}
&\begin{pmatrix}
\bbA_0 & \bbd_0 \\
\trh{\bbd_0} & c_0
\end{pmatrix} = \begin{pmatrix}
\bbM & \bbb \\
\trh{\bbb} & -\tau
\end{pmatrix}, \begin{pmatrix}
\bbA_1 & \bbd_1 \\
\trh{\bbd_1} & c_1
\end{pmatrix} = \begin{pmatrix}
\bbI_{N} & \bzero \\
\trh{\bzero} & -1
\end{pmatrix} \label{eq:qfmatrices} \\ 
&\begin{pmatrix}
 \bbA_l & \bbd_l \\
\trh{\bbd_l} & c_l
\end{pmatrix} = \begin{pmatrix}
\bbW_l & \bzero \\
\trh{\bzero} & 0
\end{pmatrix}, l=2,3,\ldots,N \label{eq:qfmatrices2}
\end{align} 
where $\bbW_l = \bbPt^{\rm R}_{l-1},l=2,3,\ldots,\frac{N+1}{2}$ and $\bbW_l = \bbPt^{\rm I}_{l-\frac{N+1}{2}},l=\frac{N+1}{2}+1,\frac{N+1}{2}+2,\ldots,N$. Comparing with \eqref{eq:quadraticform}, we have that $L=N+1$. Define the respective quadratic forms and the set as
\begin{align}
\label{eq:qform_subset}
&s_l(\bhgamma) = \trh{\begin{pmatrix}
\bhgamma    \\
-1		
\end{pmatrix}}\begin{pmatrix}
\bbA_l & \bbd_l \\
\trh{\bbd_l} & c_l
\end{pmatrix}\begin{pmatrix}
\bhgamma    \\
-1		
\end{pmatrix},l=0,1,\ldots,L-1, \\
&\Pi =  \Big\{\tr{\Big(s_0(\bhgamma), s_1(\bhgamma),\ldots,s_{L-1}(\bhgamma)\Big)}:  \bhgamma \in \calC^{N} \Big\}.
\end{align}
\begin{remark}
	\label{rem:PitoLMI}
	Let $\bbx \in \calC^{N+1}$. Since $\bhgamma \in \calC^N$, we have $ \Pi \subseteq \cQ$, where the set $\cQ$ is defined in \eqref{eq:setQ}. The matrices that comprise the quadratic forms  $q_l$ are given in \eqref{eq:qfmatrices} and \eqref{eq:qfmatrices2}.
\end{remark}
 
We are now ready to see how the primal and dual problem can yield the same optimal values. We  re-write  $\left(\cP\right)$ as
\begin{align}
\label{eq:Primal2}
{\rm Minimize}~ \calJ(\bhgamma)~{\rm s.t}~s_l(\bhgamma)=0,~l=1,\ldots,L-1
\end{align}
which  equivalently is expressed as
\begin{align}
{\rm Maximize}~&\tau    \nonumber  \\ 
&{\rm s.t}~\calJ(\bhgamma) \geq \tau, {\rm for~all~}\bhgamma~ {\rm satisfying~}s_l(\bhgamma)=0, \label{eq:primtodual_1}\\
& {\rm s.t}~s_0(\bhgamma) \geq 0,~{\rm for~all~}\bhgamma~ {\rm satisfying~}s_l(\bhgamma)=0, \label{eq:primtodual_2} \\
& {\rm s.t}~ \Pi \cap \cN = \emptyset, \label{eq:primtodual_f}
\end{align}
where $l=1,\ldots,L-1$ and the constraint $\calJ(\bhgamma) \geq \tau$ in \eqref{eq:primtodual_1} is equivalent to $s_0(\bhgamma) \geq 0$ in \eqref{eq:primtodual_2}. We obtain the final constraint after observing that the condition $ s_0(\bhgamma) \geq 0, s_l(\bhgamma)=0,l=1,\ldots,L-1$ is equivalent to \eqref{eq:primtodual_f}, where $\cN$ is defined in \eqref{eq:setN}. From Remark \ref{rem:PitoLMI}, we have that $\Pi$  is a subset of $\cQ$. Thus, $\cQ \cap \cN = \emptyset$ is a sufficient condition for $\Pi \cap \cN = \emptyset$. We, thus, replace the constraint in \eqref{eq:primtodual_f} to obtain 
\begin{align}
\label{eq:Equiprime}
{\rm Maximize}~ \tau,~{\rm s.t}~ \cQ \cap \cN = \emptyset.
\end{align}
If conditions in  Theorem \ref{appen:theorem} are satisfied then, after using \eqref{eq:qfmatrices} and \eqref{eq:qfmatrices2},  $\cQ \cap \cN = \emptyset$ is equivalent to the LMI in \eqref{eq:dual_eq} and, hence, the optimization problem in \eqref{eq:Equiprime} is nothing but the dual problem of \eqref{eq:dual_eq}. Thus, we see that solving the original primal problem is the same as solving the dual problem  and, hence,  $d^\star = p^\star$ implying  strong duality.  In the following proposition,  we show that our set $\cQ$ indeed satisfies the conditions in Theorem \ref{appen:theorem}. 
\begin{proposition}
	\label{prop:strong_duality}
$\cQ$ satisfies the conditions of Theorem \ref{appen:theorem}.
\end{proposition}
\begin{IEEEproof}
See Appendix \ref{appen:strong_duality}.
\end{IEEEproof}
\subsubsection{Computational Complexity}
We now discuss the computational complexity in obtaining  $\bhgamma_{\rm gls}$ of \eqref{eq:gammads2}. The estimator requires the coefficients $\tau^\diamond, \lambda^\diamond, \alpha^\diamond_l$ and $\beta^{\diamond}_l$ which are obtained by solving the SDP  of \eqref{eq:dual_eq}. SDPs are typically solved using interior-point algorithms, and in \cite[Chapter 11]{Nemirovski_IP}, the complexity of such methods are discussed. Applying the complexity analysis to the SDP in \eqref{eq:dual_eq}, the  resulting number of computations  is $O(N^{4.5})$.
\subsection{Normalization-based LS  (NLS)  Estimation}
One drawback with the GLS scheme is that its complexity of $O(N^{4.5})$ can be high depending upon the value of $N$. A computationally attractive alternative to the GLS scheme can be obtained by choosing $\bbT$ to be a PPT  and exploiting the time-domain equivalent of \eqref{eq:sph_base_lowdim}. 

We require that $\bhgamma$ satisfy \eqref{eq:sph_base_lowdim} whose equivalent  time-domain manifestation is given by 
\begin{align}
\label{eq:sph_base_timequi}
\abs{\bbx[i]} = \frac{1}{N}, i=0,1,\ldots,N-1,
\end{align}
where $\bbx = \trh{\bbFt}\bgamma$ and $\abs{c}$ denotes absolute value of the complex number $c$.  Thus, given an estimate of $\bgamma$, for example, the LS estimate in \eqref{eq:gamest}, we normalize its time-domain samples to have constant magnitude and transform back to the frequency-domain to obtain a refined estimate of $\bgamma$. The overall estimation procedure is shown in Table \ref{tab:NEscheme}, where two possible approaches are used  depending upon if $\bbT$ is chosen as a PPT or not. The normalization is performed by the diagonal $N \times N$ matrix $\bbX_{N}$ when $\bbT$ is chosen as a PPT and diagonal $\nc \times \nc$ matrix $\bbX_{\nc}$ when $\bbT$ is chosen otherwise.  The diagonal values of the normalization matrices are  
 \begin{align}
\bbX_{N}[i,i] = \frac{1}{N\abs{\bbxt_{\rm ls}[i]}},i=0,1,\ldots,N-1, \\
\bbX_{\nc}[i,i] = \frac{1}{\nc\abs{\bbx_{\rm ls}[i]}},i=0,1,\ldots,\nc-1.
 \end{align}
 
In Step 1 of Table \ref{tab:NEscheme}, we obtain the LS estimate which, in general, requires $N^3$ number of operations. We then transform the LS estimate to the time-domain and normalize the samples to have constant-magnitude.  When $\bbT$ is chosen as a PPT,  it suffices to only perform normalization in the $N$-dimensional space. This is because after normalization, $\bbxt_{\rm nls}$ (Step 3) satisfies \eqref{eq:sph_base_timequi} and, hence,   $\bhgamma_{\rm nls}$ (Step 4) satisfies \eqref{eq:sph_base_lowdim}. Thus,  $\bhdelta_{\rm nls} = \bbT\bhgamma_{\rm nls}$ also satisfies the phase noise geometry in the $\nc$-dimensional space when $\bbT$ is a PPT.  The  added number of computations  is mainly $2N\log(N)$ which correspond to the two $N$-point DFT operations for moving between time and frequency domain. However, when $\bbT$ is not a PPT, even after normalization, there is no guarantee that $\bhdelta_{\rm nls}$ will satisfy the phase noise geometry. To ensure that it does satisfy  when $\bbT$ is not a PPT, the normalization must be done in the $\nc$-dimensional space as shown in right half of Table \ref{tab:NEscheme}. This comes at the cost of higher computational complexity which is two $\nc$-point DFT operations. 
\begin{table}[!t]
	\caption{Normalization-based LS Estimation.} 
	\centering 
	\begin{tabular}[b]{|c|c|c|c|} 
		\toprule
		\multicolumn{2}{|c|}{When $\bbT$ is a PPT} & 		\multicolumn{2}{|c|}{When $\bbT$ is not a PPT} \\
		\midrule
		{\bf Steps} & {\bf Function} &	{\bf Steps} & {\bf Function} \\
		\midrule
		1 & $\bhgamma_{\rm ls} = \bbM^{-1}\bbb $ & 1 & $\bhgamma_{\rm ls} = \bbM^{-1}\bbb $\\
		\midrule
		2 & $\bbxt_{\rm ls} = \trh{\bbFt}\bhgamma_{\rm ls}$ & 2 & $\bhdelta_{\rm ls} = \bbT\bhgamma_{\rm ls}$ \\
		\midrule
		3 & $\bbxt_{\rm nls} = \bbX_{N}\bbxt_{\rm ls}$ & 3 & $\bbx_{\rm ls}= \trh{\bbF}\bhdelta_{\rm ls}$\\
		\midrule
		4 & $\bhgamma_{\rm nls} = \bbFt\bbxt_{\rm nls}$ &	4 &  $\bbx_{\rm nls} = \bbX_{\nc}\bbx_{\rm ls}$
		\\
		\midrule
		5 & $\bhdelta_{\rm nls} = \bbT\bhgamma_{\rm nls}$ &	5 & $\bhdelta_{\rm nls} = \bbF\bbx_{\rm nls}$ 	\\
		\midrule
		\multicolumn{2}{|c|}{Operations $\approx N^3+ 2N\log(N)$} & 		\multicolumn{2}{|c|}{Operations $\approx  N^3+ 2\nc\log(\nc)$} \\
		\bottomrule
	\end{tabular}
	\label{tab:NEscheme} 
\end{table}
\section{Numerical Results}
\label{sec:NR}
We now present numerical results of the proposed phase noise estimation schemes and compare them with some of the state-of-the-art scattered pilot-based phase noise estimation  schemes. In particular, we compare our proposed GLS and NLS scheme with the ULS scheme of \cite{1033877} and the CPE-based interpolation schemes of \cite{5508318} and \cite{5068965}. 

The  system parameters set for the simulations are as follows: The number of subcarriers $\nc = 512$; subcarrier spacing $f_{\rm sub} = 15$ kHz; bandwidth is $7.7$ MHz. The percentage of scattered pilot subcarriers is set to $8\%$ and symbol constellation is $16$-QAM.  The channel is Rayleigh fading  with  four exponentially decaying taps, and coherence bandwidth is set to $800$ kHz. We use a $1/2$-rate convolutional  encoder $\left[133, 171\right]$ with  constraint length of $7$. For decoding, we use a soft-decision Viterbi decoder of decoding depth equal to five times the constraint length. Phase noise process used in the simulations is the Wiener process which models well  free-running oscillators.  We denote  $f_{\rm 3dB}$ as the phase noise $3$-dB bandwidth, and the quantity $\varrho  = \frac{f_{\rm 3dB}}{f_{\rm sub}}$ is a measure of how fast or slow the phase noise varies within an OFDM symbol. A low value of $\rho$ indicates a slow-varying phase noise process while a larger value indicates a fast-varying one. 

The  phase noise estimation schemes of this paper require knowledge of the channel. This knowledge is  acquired by estimating the channel.   We refer the reader to \cite{1677918, 5508318,  4567677, 4568453, 6297481, 6868950} for some of the state-of-the-art methods on channel estimation in the presence of phase noise and  frequency offset.  In this paper, we use the channel estimator of \cite{5508318} which is computationally attractive compared to other schemes and at the same time  takes into account the effect of phase noise during the estimation process.
\subsection{CPE-based Interpolation Scheme (CIS) of \cite{5508318} and \cite{5068965}}
We now briefly summarize the interpolation schemes of \cite{5508318} and \cite{5068965}. The goal is to develop a non-iterative scheme for phase noise estimation for data OFDM symbols. Such a phase noise estimate is obtained as follows: The CPE of the current and next OFDM symbol are estimated using scattered pilot subcarriers. The average value of  phase noise in the current and next OFDM symbol is then obtained by taking the angle of the obtained CPE estimates. The mean phase noise values are then interpolated to obtain the entire phase noise realization between the mid-points of the current and next OFDM symbols. A linear interpolator is used in both \cite{5508318} and \cite{5068965}. In fact, it is shown in \cite{5508318} that for slow-varying phase noise processes, the optimal interpolator, in terms of minimum mean square error, is the linear interpolator. The CIS schemes are simple and computationally very  attractive. However, for moderately or fast-varying phase noise, we can expect an inferior performance which is verified by the numerical results. 
\subsection{Discussion}
\begin{figure}[!t]
	\begin{center}
			\psfrag{SNR}{{\footnotesize SNR~{\bf [dB]}}}
		\includegraphics[height=0.3\textheight, width=0.47\textwidth]{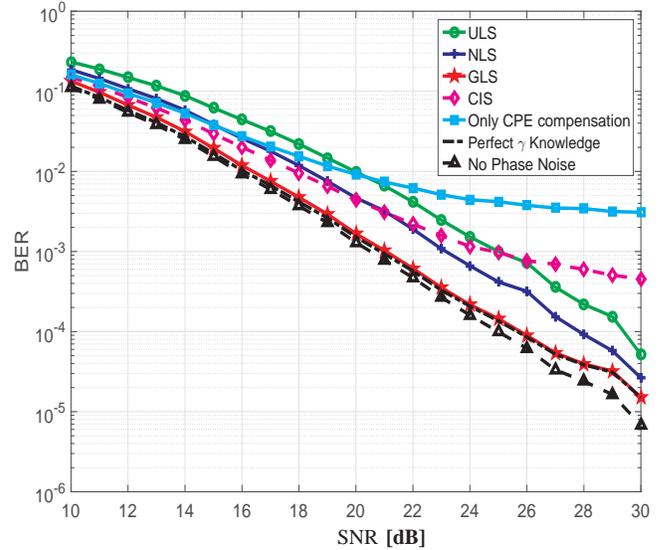}
		\caption{Comparison of  average coded BER vs.~SNR for the  proposed schemes with $N=8$ and $\varrho = 0.02$. The transformation matrix used is $\bbT_{\rm pc}$ of \eqref{eq:PCPPT}.}
		\label{fig:coBER_vs_snr}
	\end{center}
\end{figure}

Figure \ref{fig:coBER_vs_snr} shows coded BER  performance of the proposed phase noise estimation schemes. The ideal performance that can be achieved is shown by the triangle-marker  dashed curve which corresponds to the case of zero   phase noise. The squared-marker curve represents the case where only CPE compensation is performed. This method works well only for extremely slow-varying phase noise processes. As seen from the figure, the best performance is achieved by the  GLS scheme and is close to the ideal performance. It also outperforms the CIS schemes of \cite{5508318} and \cite{5068965} as expected. The GLS scheme constraints the LS estimator to adhere to the phase noise geometry. As seen in the figure, the ULS scheme, which is the unconstrained LS estimator, has an inferior performance compared to its constrained GLS counterpart. The NLS scheme is a suboptimal solution that also achieves the same objective of delivering an estimate that satisfies the phase noise geometry. As expected, the NLS scheme has a better performance compared to the LS scheme. 

\begin{figure*}[!t]
\centerline{
		\psfrag{Prob}[c]{{\footnotesize Probability density}}
		\psfrag{Err}[c]{{\footnotesize $\norm{\bhdelta-\bbdelta}^2$  {\bf [\%]}}}
		\subfloat[SNR equal to $30$-dB.]{
		\includegraphics[height=0.3\textheight, width=0.47\textwidth]{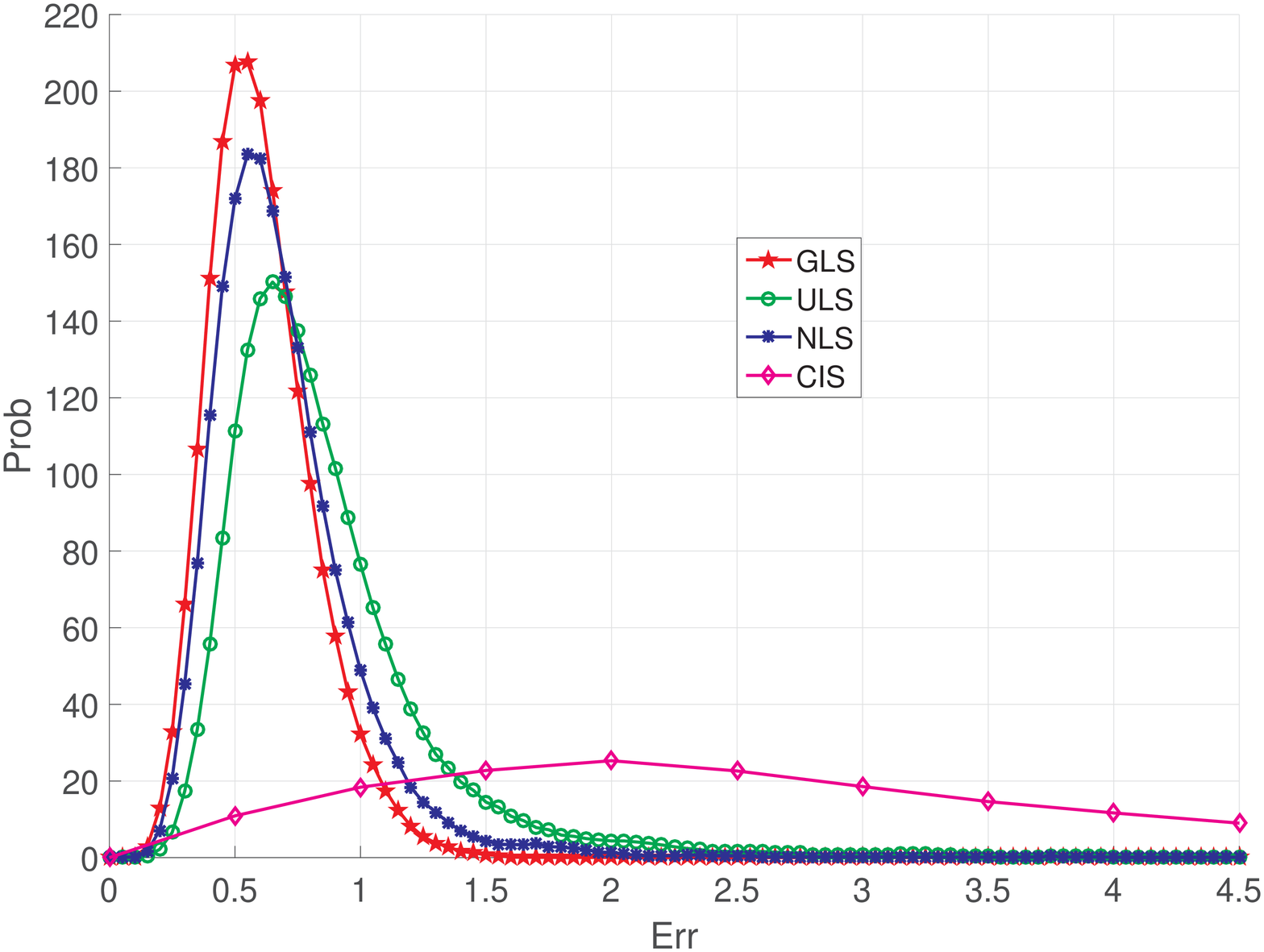}
		\label{fig:Errpdf_snr_ppt_30}}
\hfil
		\psfrag{Prob}[c]{{\footnotesize Probability density}}
		\psfrag{Err}[c]{{\footnotesize  $\norm{\bhdelta-\bbdelta}^2$  {\bf [\%]}}}
		\subfloat[SNR equal to $10$-dB.]{
		\includegraphics[height=0.3\textheight, width=0.47\textwidth]{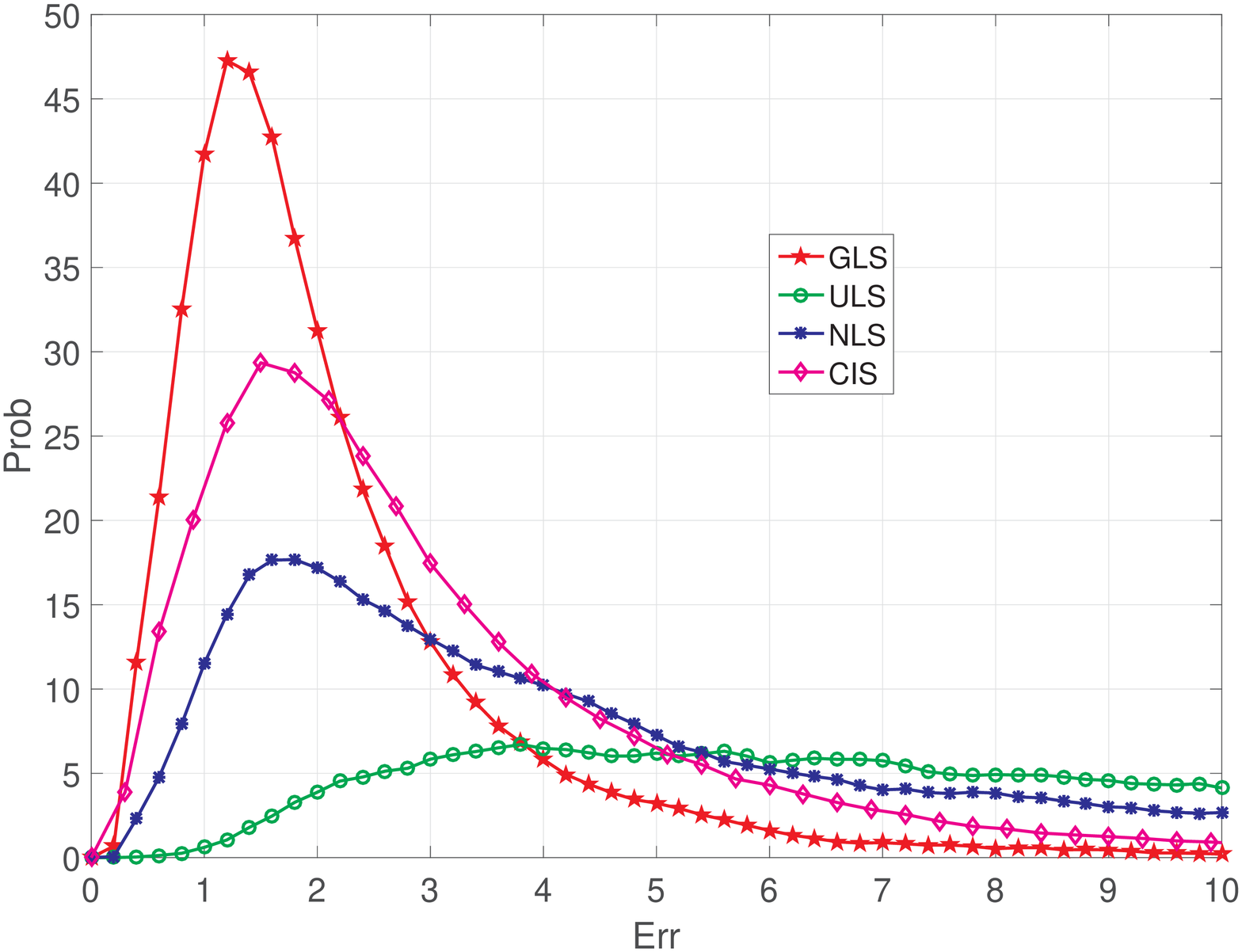}
		\label{fig:Errpdf_snr_ppt_10}}}
\caption{Empirical PDF of  $\norm{\bhdelta-\bbdelta}^2$ for the proposed schemes with $N=8$  and $\varrho = 0.02$. The transformation matrix used is $\bbT_{\rm pc}$ of \eqref{eq:PCPPT}.}
\label{fig:Errpdf}
\end{figure*}

The BER  performance of the phase noise estimation schemes can be explained by examining the PDF of  $\norm{\bhdelta-\bbdelta}^2$, where $\bhdelta$ is our estimate of the true value of $\bbdelta$. In Figs.~\ref{fig:Errpdf_snr_ppt_30} and \ref{fig:Errpdf_snr_ppt_10}, we plot the empirical PDF of $\norm{\bhdelta-\bbdelta}^2$ for SNR of $30$-dB and $10$-dB, respectively. From Fig.~\ref{fig:Errpdf_snr_ppt_30}, we see that the GLS scheme exhibits thinner tails in the PDF  compared to all other schemes. The thicker tails seen, for example, in the ULS scheme results in a higher BER  as verified in Fig.~\ref{fig:coBER_vs_snr} at SNR equal to $30$-dB. In  Fig.~\ref{fig:Errpdf_snr_ppt_10}, at the lower SNR of $10$-dB, for all schemes, we see that the PDF of the phase noise estimation error is spread over a large range of values, thereby, resulting in a much higher BER.

\begin{figure}[!t]
	\psfrag{Rho}{{\footnotesize $\varrho$}}
	\begin{center}
		\includegraphics[height=0.3\textheight, width=0.47\textwidth]{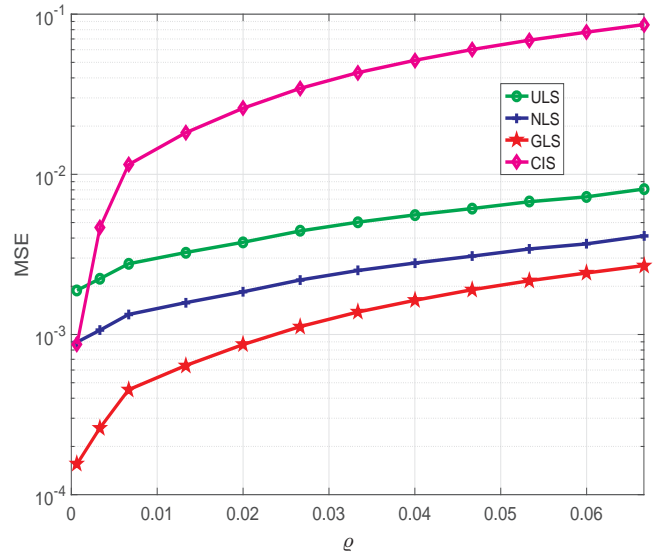}
		\caption{Comparison of  MSE of $\bhgamma$ vs.~$\varrho$ for the  proposed schemes with  $N=8$.  The transformation matrix used is $\bbT_{\rm pc}$ of \eqref{eq:PCPPT}. The SNR is set at $30$-dB.}
		\label{fig:MSE_vs_beta}
	\end{center}
\end{figure}

A moderate value of $\varrho = 0.02$ was used in the simulation results shown in Figs.~\ref{fig:coBER_vs_snr}, \ref{fig:Errpdf_snr_ppt_30} and \ref{fig:Errpdf_snr_ppt_10}. It is of practical interest to see how well the proposed algorithms perform over the practical range of values of $\varrho$. This is demonstrated in Fig.~\ref{fig:MSE_vs_beta}, where we plot the  mean-square-error (MSE) of $\bhgamma$,\ie, ${\rm E}\left[\norm{\bhgamma -\bgamma}^2\right]$ as a function of $\varrho$. A small value of $\varrho$ indicates a slow-varying phase noise process in comparison with the OFDM symbol duration and vice-versa.  As expected and verified in the figure, MSE of $\bhgamma$, in general, increases with $\varrho$. The best performance is obtained by the GLS scheme with  CIS  performing the poorest. This is easily seen since the CIS scheme obtains the entire phase noise realization using a linear interpolator. As the value of $\varrho$ increases, the phase noise realization is more fast-varying in nature, and a simple linear interpolator does a poor job of approximation.

\begin{figure}[!t]
	\begin{center}
		\psfrag{SNR}{{\footnotesize SNR~{\bf [dB]}}}
		\includegraphics[height=0.3\textheight, width=0.47\textwidth]{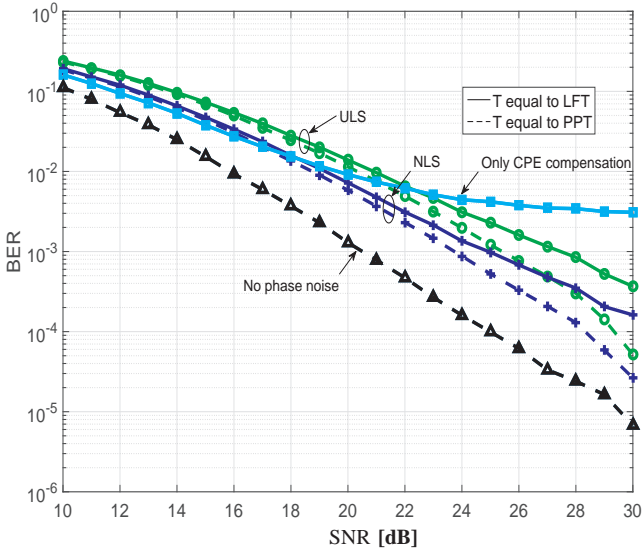}
		\caption{Effect of transformation matrix $\bbT$ on average coded-BER when $\bbT = \bbT_{\rm pc}$ of \eqref{eq:PCPPT} is compared with $\bbT = \bbL$ of \eqref{eq:selection_mat}. The value of  $N=8$ and $\varrho = 0.02$.}
		\label{fig:Ber_compare_vs_snr}
	\end{center}
\end{figure}

\begin{figure*}[!t]
\centerline{
		\psfrag{Prob}[c]{{\footnotesize Probability density}}
		\psfrag{Err}[c]{{\footnotesize  $\norm{\bhdelta-\bbdelta}^2$  {\bf [\%]}}}
		\subfloat[SNR equal to $30$-dB. Observe that the PDF has thicker tails when $\bbT$ is set to LFT, thus, exhibiting higher BER in Fig.~\ref{fig:Ber_compare_vs_snr} at $30$-dB SNR.]{\includegraphics[height=0.3\textheight, width=0.47\textwidth]{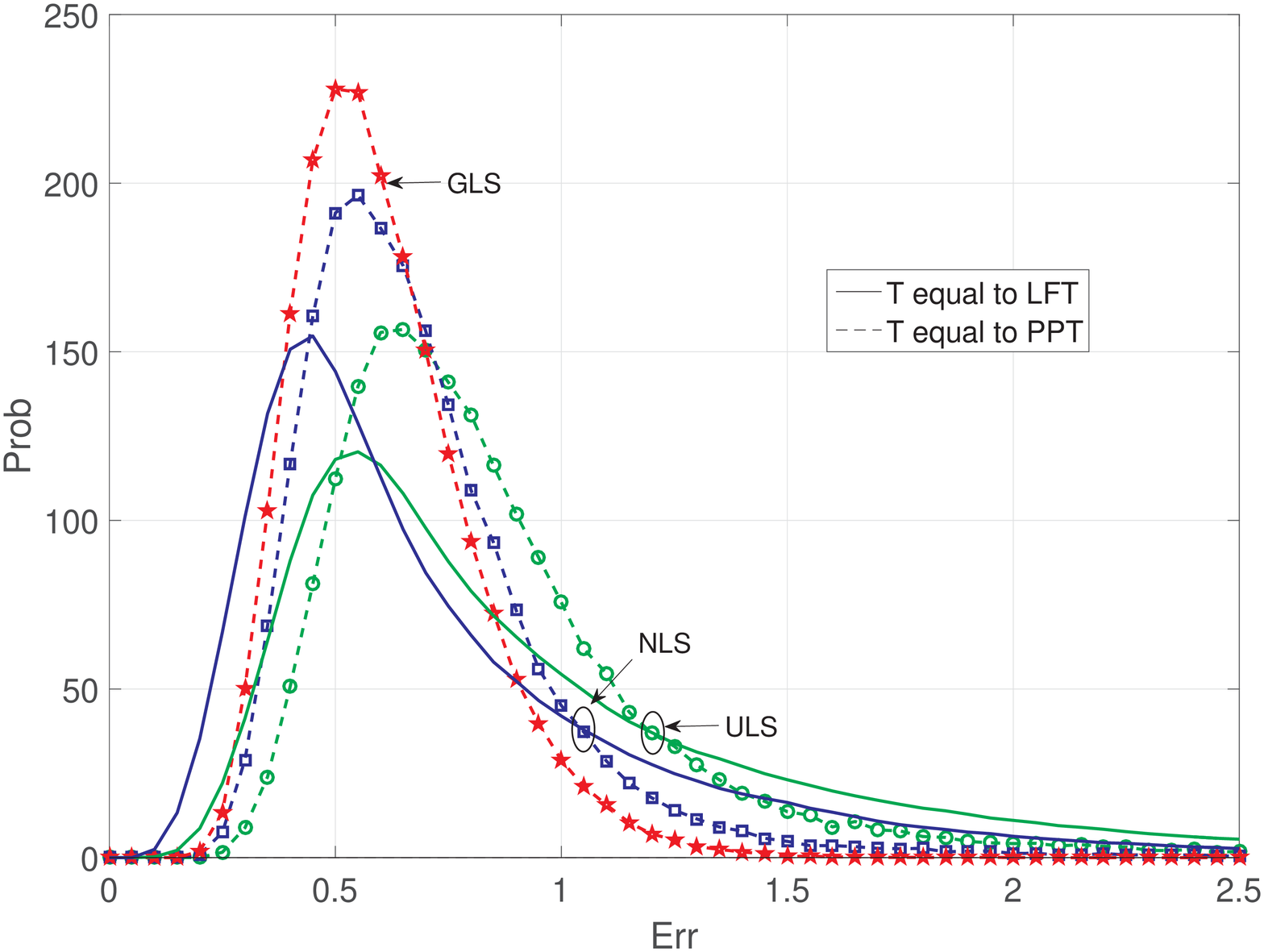}
		\label{fig:Errpdf_snr_30}}
\hfil
		\psfrag{Prob}[c]{{\footnotesize Probability density}}
		\psfrag{Err}[c]{{\footnotesize  $\norm{\bhdelta-\bbdelta}^2$  {\bf [\%]}}}
		\subfloat[SNR equal to  $10$-dB.  Observe that for ULS and NLS,  there is no dependence of  the PDF tails  on the choice of $\bbT$, thus, exhibiting similar  BER values in  Fig.~\ref{fig:Ber_compare_vs_snr} at $10$-dB SNR.]{\includegraphics[height=0.3\textheight, width=0.47\textwidth]{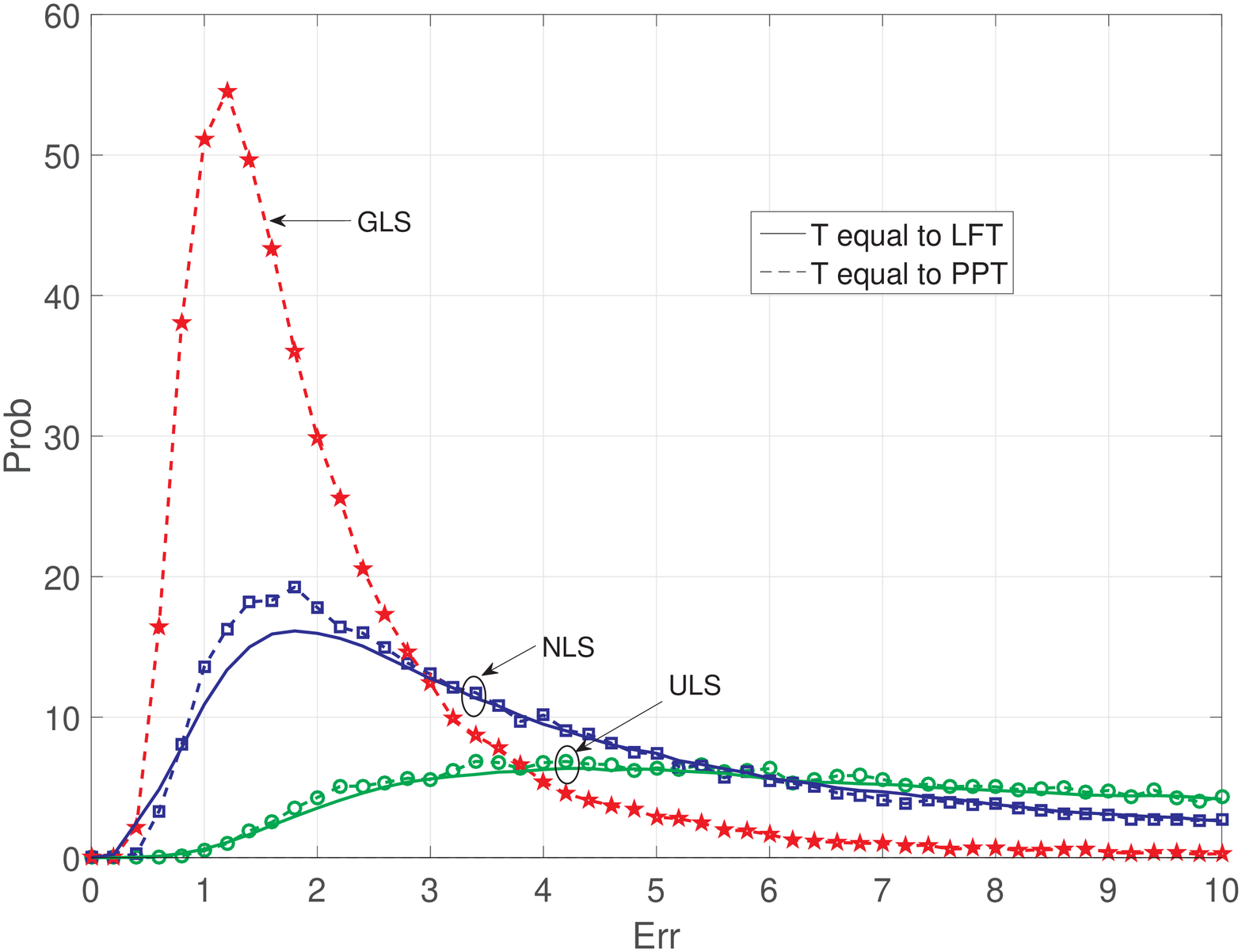}
	\label{fig:Errpdf_snr_10}}}
	\caption{Effect of  $\bbT$ on the empirical PDF of  $\norm{\bhdelta-\bbdelta}^2$ for the proposed schemes. $\bbT_{\rm pc}$ of \eqref{eq:PCPPT} is used as the PPT and $\bbL$ of \eqref{eq:selection_mat} is the LFT. The number of estimated components is $N=8$. The value of $\varrho = 0.02$. The  GLS is also plotted for comparison. It is always implemented with $\bbT$ set to a PPT.}
\end{figure*}

We now compare the effect of the transformation matrix $\bbT$ on the proposed phase noise estimation schemes. 
Figure \ref{fig:Ber_compare_vs_snr} shows the average coded-BER for the ULS and NLS schemes with $\bbT$  set to $\bbT_{\rm pc}$ of \eqref{eq:PCPPT} and with $\bbT=\bbL$ of \eqref{eq:selection_mat}. From the figure we see that for $\bbT$ equal to PPT, the ULS and NLS schemes yield a lower average BER compared to the case when $\bbT$ is set as LFT,  especially at high SNRs. We can again explain this behavior by examining the PDF of $\norm{\bhdelta-\bbdelta}^2$ which is shown in Fig.~\ref{fig:Errpdf_snr_30}, where SNR is set to $30$-dB. From the figure, we see that when $\bbT$ is equal to the LFT of  \eqref{eq:selection_mat}, the empirical PDF, of both  ULS and NLS, exhibits thicker tails compared to the curves $\bbT$ equal to PPT. Also plotted in the figure  is the GLS scheme. Note that for GLS $\bbT$ is set to $\bbT_{\rm pc}$ of \eqref{eq:PCPPT}. These thicker tails eventually cause higher BER as observed in Fig.~\ref{fig:Ber_compare_vs_snr} at SNR equal to $30$-dB.  Figure \ref{fig:Errpdf_snr_10} shows the empirical PDF at SNR equal to $10$-dB. As can be seen, for any choice of $\bbT$, the ULS and NLS exhibit similar behavior especially at the tails of the PDF. Thus, we can expect  similar  BER as evidenced in Fig.~\ref{fig:Ber_compare_vs_snr} at SNR of $10$-dB. 
 
\begin{figure*}[!t]
\centerline{\psfrag{Amp}[c]{{\footnotesize Amplitude  {\bf [radians]}}}
				\psfrag{realization}[c]{{\footnotesize Samples}}
		\subfloat[$\bbT$ set to LFT of  \eqref{eq:selection_mat}.]{\includegraphics[height=0.3\textheight, width=0.47\textwidth]{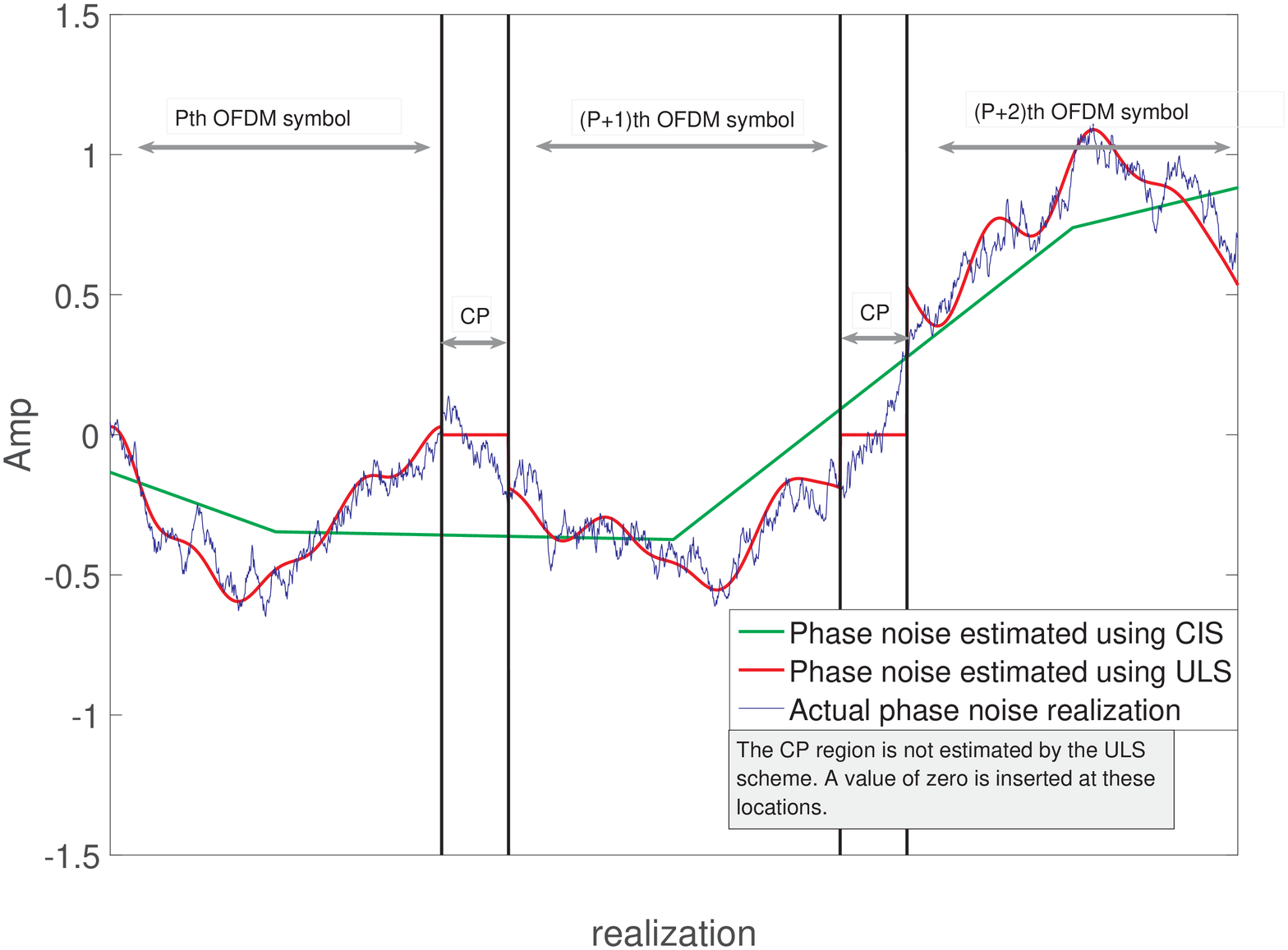}
		\label{fig:pn_realization_lft}}
\hfil
		\psfrag{Amp}[c]{{\footnotesize Amplitude  {\bf [radians]}}}
		\psfrag{realization}[c]{{\footnotesize Samples}}
		\subfloat[$\bbT$ set to PPT of \eqref{eq:PCPPT}.]{\includegraphics[height=0.3\textheight, width=0.47\textwidth]{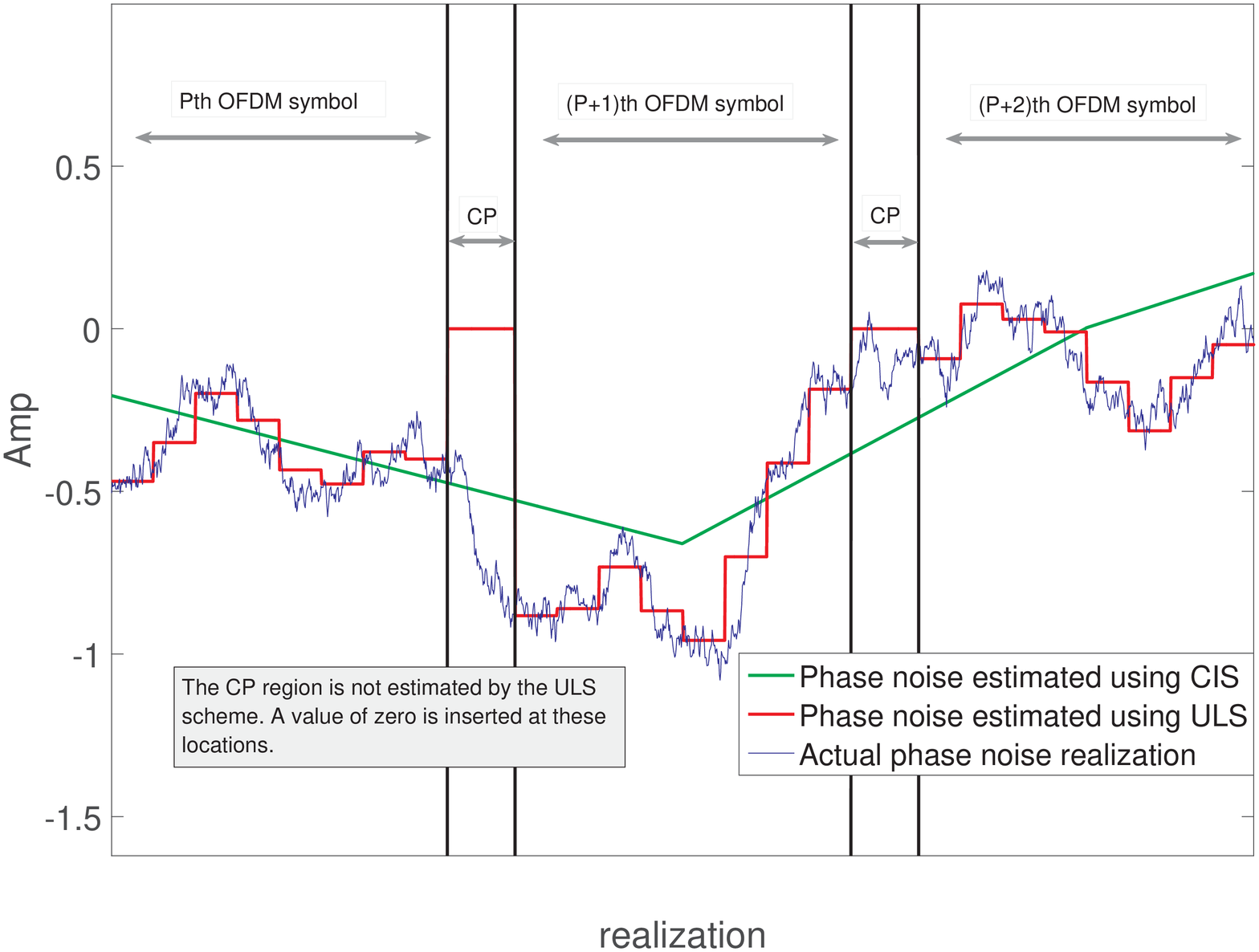}
				\label{fig:pn_realization_ppt}}}
		\caption{Comparison of the estimated phase noise realization with the actual phase noise realization. The value of  $N=8$ and $\varrho = 0.02$ with $30$-dB SNR.}			
\end{figure*}

The effect of the transformation matrix $\bbT$ can also be visualized by looking at the estimated phase noise realization. We illustrate this effect, for example, using the ULS scheme.  Figures \ref{fig:pn_realization_lft} and \ref{fig:pn_realization_ppt} show, respectively, the estimated phase noise realization when $\bbT$ is set as a LFT and a PPT. For comparison, we also plot the estimated phase noise realization using the CIS scheme. From  Fig.~\ref{fig:pn_realization_lft}, we observe that the LFT matrix $\bbL$ of \eqref{eq:selection_mat} allows only for smooth approximation of the true phase noise realization. This is because the model in \eqref{eq:selection_mat}  estimates $N$ low-frequency components.  For example, in the figure,  $N=8$ which implies eight low-frequency components are estimated. On the other hand, in Fig.~\ref{fig:pn_realization_ppt}, we observe that when $\bbT$ is set to the PPT of   \eqref{eq:PCPPT},  a  piece-wise approximation of the phase noise realization is obtained. This effect arises  because the interpolation matrix in \eqref{eq:PCPPT} is a piece-wise constant interpolator.  In both the figures, we observe that, using the CIS scheme, the estimated phase noise realization is  a linear approximation of the true phase noise realization. As seen in the figure, for the set value of $\varrho=0.02$ which results in a moderately-varying phase noise process, the linear approximation is a poor estimate.  
\section{Conclusion}
This paper presents scattered pilot-based phase noise estimation schemes for an OFDM radio link corrupted by phase noise. Pilot-based estimation schemes are attractive for delay sensitive wireless systems when compared to decision-feedback schemes which can incur significant computational load and, hence, delay onto the  receiver. This paper builds upon earlier work wherein, using the least-squares principle,  phase noise is estimated from scattered pilot subcarriers. It is shown that such an estimator suffers from  amplitude and phase estimation errors which arises due to receiver noise, estimation from limited scattered pilot subcarriers and estimation using a dimensionality reduction model. We  empirically show that the phase estimation error is small and the critical factor is the amplitude estimation error. To eliminate the amplitude estimation error, the least-squares estimate  is enforced to satisfy the so-called phase noise spectral geometry. Numerical results demonstrate superior bit-error-rate and phase noise estimation error performance for the estimator  that abides by this geometry.
\appendices
\section{Proof of Proposition \ref{prop:strong_duality}}
\label{appen:strong_duality}
The proof follows on similar lines as  in \cite{7366606}. From Theorem \ref{appen:theorem}, we  need to prove the following:
\begin{itemize}
\item[P1.] The set $\bbQt$ satisfies the regularity conditions,\ie, its conic hull spans the entire $\cR^{L-1}$, where $L=N+1$. 
\item[P2.] $\cQ \cap \cN =  \emptyset$ implies ${\rm cov}\left(\cQ\right) \cap \cN =  \emptyset$.
\end{itemize}
We begin with P1. 
\subsection{Proof of P1}
\label{sec:P1}
 The set $\cQt$ is described by the quadratic forms of \eqref{eq:qfmatrices} and \eqref{eq:qfmatrices2},\ie,
\begin{align}
\label{eq:qtfmatrices}
q_1(\bbx) = \trh{\bbx}\begin{pmatrix}
\bbI_{N} & \bzero \\
\trh{\bzero} & -1
\end{pmatrix}\bbx, q_l(\bbx) = \trh{\bbx}\begin{pmatrix}
\bbW_l & \bzero \\
\trh{\bzero} & 0
\end{pmatrix}\bbx,
\end{align}
where $\bbW_l = \bbPt^{\rm R}_{l-1},l=2,3,\ldots,\frac{N+1}{2}$ and $\bbW_l = \bbPt^{\rm I}_{l-\frac{N+1}{2}},l=\frac{N+1}{2}+1,\frac{N+1}{2}+2,\ldots,N$.   Let  $\{\bbft_i\}_{i=1}^{N}$ denote  column vectors of the $N\times N$ DFT matrix $\bbFt$. First, we note that the permutation matrix $\bbPt_l$ is circulant and, hence, diagonalizable by $\bbFt$. The eigenvalues of $\bbPt_l$ are given by $\{e^{\jmath\frac{2\pi nl}{N}}\}_{n=0}^{N-1}$ and, thus, the eigenvalues of $\bbPt^R_l$ and $\bbPt^I_l$ are $\{\cos(\frac{2\pi nl}{N})\}_{n=0}^{N-1}$ and $\{\sin(\frac{2\pi nl}{N})\}_{n=0}^{N-1}$, respectively. We are now ready to prove the regularity condition. 

Choose $\bbx_i = \tr{\left[\tr{\bbft_i}~0\right]}, i=1,2,\ldots,N$ and $\bbx_{N+1} = \tr{\left[\tr{\bzero}~\sqrt{N}\right]}$,\ie, we choose $M = N+1$ points. We note that $M > L-1$ since $L = N+1$. Making use of the eigenvalues of $\bbPt^R_l$ and $\bbPt^I_l$, the points $\bbq(\bbx_i)$ and, hence, the matrix $\bbQ$ of \eqref{eq:Qmat} is given by
\begin{align}
\label{eq:Qmatex}
\bbQ = \mbox{\tiny$
	\begin{pmatrix}
	\mathstyle  1 & \mathstyle 1 & \mathstyle 1 & \mathstyle \ldots  & \mathstyle 1 & \mathstyle -N \\
	\mathstyle  1 & \mathstyle \cos(\frac{2\pi}{N}) & \mathstyle \cos(\frac{4\pi}{N}) &\mathstyle \ldots  & \mathstyle \cos(\frac{2\pi(N-1)}{N}) & \mathstyle 0 \\
	\mathstyle \vdots &  \mathstyle \vdots & \mathstyle \vdots & \mathstyle \vdots & \mathstyle \vdots & \mathstyle \vdots\\
	\mathstyle 1 & \mathstyle \cos(\frac{2\pi(N-1)}{2N}) & \mathstyle \cos(\frac{4\pi(N-1)}{2N}) & \mathstyle \ldots  & \mathstyle \cos(\frac{2\pi(N-1)(N-1)}{2N}) & \mathstyle \vdots\\
	\mathstyle 0 & \mathstyle \sin(\frac{2\pi}{N}) &  \mathstyle\sin(\frac{4\pi}{N}) & \mathstyle \ldots  & \mathstyle  \sin(\frac{2\pi(N-1)}{N}) & \mathstyle \vdots\\
	\mathstyle \vdots &  \mathstyle \vdots & \mathstyle \vdots & \mathstyle \vdots & \mathstyle \vdots & \mathstyle \vdots\\
	\mathstyle	0 & \mathstyle \sin(\frac{2\pi(N-1)}{2N}) & \mathstyle \sin(\frac{4\pi(N-1)}{2N}) & \mathstyle \ldots  & \mathstyle \sin(\frac{2\pi(N-1)(N-1)}{2N}) & \mathstyle 0  
	\end{pmatrix}$}.
\end{align}
From \eqref{eq:Qmatex},  we note that ${\rm rank}\left(\bbQ\right) = N$ since the rows form an orthogonal basis. Choose constants $\{p_i\}_{i=1}^{M} = 1$. Then $\sum_{i=1}^{M=N+1}p_i\bbq(\bbx_i) = \bzero$ since the elements of each row  sum to a value of zero. This completes the proof.  
\subsection{Proof of P2}
The set $\cQ$ is defined in \eqref{eq:setQ} and described by the quadratic forms $q_l(\bbx), l=0,1,\ldots N$, where $q_l(\bbx), l>0$ is given in \eqref{eq:qtfmatrices}. The quadratic form $q_0(\bbx)$ takes the form
\begin{align}
\label{eq:q0}
q_0(\bbx) =  \trh{\bbx}\begin{pmatrix}
	\bbM & \bbb \\
	\trh{\bbb} & -\tau
	\end{pmatrix}\bbx.
	\end{align}
Consider the set
\begin{align}
\label{eq:Qn}
\cQ_N &= \Big\{\tr{\Big(q_0(\bbx),~q_1(\bbx),\ldots,q_{N}(\bbx)\Big)}:  ||\bbx||_2 = 1, \bbx \in \calC^{N+1} \Big\}.
\end{align}
It is related to $\cQ$ by \cite{dines, brickman}
\begin{align}
\label{eq:QnrQ}
\cQ = \left\{ t\bby~\Big{|}~ t\geq 0,~\bby \in \cQ_N\right\}.
\end{align}  
Let ${\rm cov}\left(\cQ_N\right)$ denote the convex hull of $\cQ_N$.  We  define  
\begin{align}
\label{eq:conQ}
{\rm con}\left(\cQ\right) = \left\{ t\bby~\Big{|}~ t\geq 0,~\bby \in {\rm cov}\left(\cQ_N\right)\right\}.
\end{align}
First, we observe that $ \cQ \subseteq {\rm con}\left(\cQ\right)$. Secondly, ${\rm con}\left(\cQ\right)$ is a convex set since it is defined in terms of the convex set $ {\rm cov}\left(\cQ_N\right)$. We, thus, have
\begin{align}
\label{eq:consupcov}
{\rm cov}\left(\cQ\right) \subseteq  {\rm con}\left(\cQ\right),
\end{align}
since ${\rm cov}\left(\cQ\right)$ is the convex hull of $\cQ$ and by definition is the smallest convex set enclosing $\cQ$. With these facts in place, we have the following relation:
\begin{itemize}
\item[R1.] $\cQ \cap \cN = \emptyset \equiv \cQ_N \cap \cN = \emptyset$. 
\item[R2.] ${\rm cov}\left(\cQ_N\right) \cap \cN = \emptyset \implies {\rm con}\left(\cQ\right) \cap \cN = \emptyset \implies {\rm cov}\left(\cQ\right) \cap \cN = \emptyset$,
\end{itemize}
where $\equiv$ denotes equivalence and $\implies$ denotes implication. The equivalence in R1  follows from  \eqref{eq:Qn} and \eqref{eq:QnrQ}. The implication in R2 follows from  \eqref{eq:conQ}. We, thus, see that if $\cQ_N \cap \cN = \emptyset \implies {\rm cov}\left(\cQ_N\right) \cap \cN = \emptyset$ then, after combining R1 and R2, we have the required result.  We now show that this is indeed the case. 
\begin{remark}
	\label{rem:q1cond}
	For unit-norm $\bbx$, 	$q_1(\bbx) = 0$ only at $\bbx = \tr{\left[\sqrt{0.5}\tr{\bbxt} \sqrt{0.5}z\right]}$, 
	where  $\norm{\bbxt}_2=1$ and $|z| = 1$.
\end{remark}
\begin{proposition}
	\label{prop:Qn}
For unit-norm $\bbx$, $q_l(\bbx) = 0$ for all $l > 1$  at 
	\begin{align}
	\bbx = \tr{\left[\sqrt{a}\tr{\bbxt} \sqrt{b}z\right]}, 
\bbxt = \bbF\Sigma\bbvt, \tr{\bbvt}\bbvt = 1, |z|=1, \label{eq:qallcond_xt}
	\end{align}
where $\bbvt_i = \frac{1}{\sqrt{N}}$, $a \geq 0, b \geq 0$, $a+b = 1$ and $\Sigma$ can be any unitary-diagonal matrix. 	
\end{proposition}
\begin{IEEEproof}
	Write $\bbx = \tr{\left[\sqrt{a}\tr{\bbxt} \sqrt{b}z\right]}$. Since $\bbx$ should be of unit-norm, we have $\norm{\bbxt}_2 = 1$, $a \geq 0, b \geq 0$, $a+b = 1$ and $|z| = 1$. Using \eqref{eq:qtfmatrices}, the condition $q_l(\bbx) = 0, l > 1$ results in 
	\begin{align}
		\label{eq:matAnull}
		&\trh{\bbxt}\bbW_l\bbxt =  0,  \\
		&\trh{\bbxt}\bbFt\bbDt_l\trh{\bbFt}\bbxt = 0, \\
		&\trh{\bby}\bbDt_l\bby = \tr{\bbd_l}\bbv = 0, \label{eq:finalcond}
	\end{align}
	where $ \bby = \trh{\bbFt}\bbxt$ with components denoted by $\bby_i$ and  $\bbv = \tr{\left[|\bby_0|^2~|\bby_1|^2 \ldots |\bby_{N-1}|^2    \right]}$. In the above equation, we used the fact that $\bbW_l$ is diagonalizable with the DFT matrix whose  eigenvalues are contained in the  diagonal matrix $\bbDt_l$ and in the vector $\bbd_i$.  Combining \eqref{eq:finalcond} for all $l \geq 2$, we have
	\begin{align}
		\label{eq:Qmatnew}
		\mbox{\tiny$\begin{pmatrix}
			1 & \cos(\frac{2\pi}{N}) & \cos(\frac{4\pi}{N}) & \ldots  & \cos(\frac{2\pi(N-1)}{N}) \\
			\vdots &  \vdots & \vdots & \vdots & \vdots \\
			1 & \cos(\frac{2\pi(N-1)}{2N}) & \cos(\frac{4\pi(N-1)}{2N}) & \ldots  & \cos(\frac{2\pi(N-1)(N-1)}{2N}) \\
			0 & \sin(\frac{2\pi}{N}) & \sin(\frac{4\pi}{N}) & \ldots  & \sin(\frac{2\pi(N-1)}{N}) \\
			\vdots &  \vdots & \vdots & \vdots & \vdots \\
			0 & \sin(\frac{2\pi(N-1)}{2N}) & \sin(\frac{4\pi(N-1)}{2N}) & \ldots  & \sin(\frac{2\pi(N-1)(N-1)}{2N}) \\
			\end{pmatrix}$}\bbv = \bzero,
	\end{align} 
	where we require that $\bbv \succeq 0$ and $\norm{\bbv}_1=1$ because $\norm{\bbxt}_2=1$. It can be easily seen that the  above matrix has a non-zero null space of rank equal to one. The vector describing this space (and satisfying  $\bbv \succeq 0$,  $\norm{\bbv}_1=1$) is given by 
	\begin{align}
		\label{eq:vvec}
		\bbv = \frac{1}{N}\bone,
	\end{align}
	where $\bone$ denotes  N-dimensional vector of ones. Define $\bbvt$ as the vector with  elements $\bbvt_i = \sqrt{\bbv_i}$. Thus, at $\bbxt = \bbFt\Sigma\bbvt$, where $\Sigma$ can be any unitary-diagonal matrix,  $\trh{\bbxt}\bbW_l\bbxt =  0$, for all $l\geq 2$. 
\end{IEEEproof}
\begin{proposition}
	\label{prop:infimumcond}
For any $a \geq 0$ and $b \geq 0$, such that $a > b$ and $a+b = 1$, we have
\begin{align}
&\rm{infimum}\left(\trh{\bbxt}\bbA\bbxt  + 2\rm{Real}\left(\trh{\bbxt}\bbc \right)\right) \leq \nonumber \\ &~~~~\rm{infimum}\left(\trh{\bbxt}\bbA\bbxt + \sqrt{\frac{b}{a}}2\rm{Real}\left(\trh{\bbxt}\bbc\right)\right)~,
\end{align}
where  $\bbA \succ 0$, $\bbc$ is any complex vector and the infimum is taken over all $\bbxt$ satisfying \eqref{eq:qallcond_xt}.
\end{proposition}
\begin{IEEEproof}
	First, we note  there exists an $\bbxt$ satisfying \eqref{eq:qallcond_xt} such that ${\rm Real}\left(\trh{\bbxt}\bbc \right) \leq 0$. For example, from  \eqref{eq:qallcond_xt},  the components of the  row vector $\trh{\bbvt}\trh{\Sigma}$ take the form $\frac{e^{-\jmath\phi_l}}{\sqrt{N}}$, where $e^{\jmath\phi_l}$ are the diagonal values of diagonal $\Sigma$ matrix. Since $\Sigma$ can be any unitary-diagonal matrix, set  $\phi_l =  \angle{\left(\trh{\bbFt}\bbc\right)_l - \pi}$, where $\angle{x}$ denotes angle of the complex number $x$. Thus, ${\rm Real}\left(\trh{\bbvt}\trh{\Sigma}\trh{\bbFt}\bbc \right) = -\norm{\trh{\bbFt}\bbc}_1 \leq 0$. Thus, we have
	\begin{align}
	\rm{infimum}\left(\trh{\bbxt}\bbA\bbxt  + 2\rm{Real}\left(\trh{\bbxt}\bbc \right)\right) &= \eta - \epsilon \\
	\rm{infimum}\left(\trh{\bbxt}\bbA\bbxt  + \sqrt{\frac{b}{a}}2\rm{Real}\left(\trh{\bbxt}\bbc \right)\right) &= \eta - \sqrt{\frac{b}{a}}\epsilon,
	\end{align}
where $\epsilon \geq 0$ and $\eta$ is the minimum eigenvalue of $\bbA$. The result now follows since $a > b$. 	
	\end{IEEEproof}

Let ${\rm con}\left(\cQ_N\right)$ denote the conic hull of $\cQ_N$. We now have the following 	proposition: 
\begin{proposition}
	\label{prop:conichullq}
Let $\tau \leq \rm{infimum}\left(\trh{\bbxt}\bbM\bbxt  + 2\rm{Real}\left(\trh{\bbxt}\bbb z\right)\right)$. 	The point $\tr{\left[0, 1, 0\ldots 0 \right]} \notin {\rm con}\left(\cQ_N\right)$.
\end{proposition}
\begin{IEEEproof}
 If	 $\tr{\left[0, 1, 0\ldots 0 \right]} \in {\rm con}\left(\cQ_N\right)$ then there must exist  $\tr{\left[0, t, 0\ldots 0 \right]} \in \cQ_N$ for some $t > 0$ \cite{dines}. We show that this is impossible. From  Proposition \ref{prop:Qn}, we have $q_l(\bbx) = 0, l > 1$ for $\bbx$ of \eqref{eq:qallcond_xt}. At such an $\bbx$, $q_1(\bbx) = a - b$ and, since we require  $\tr{\left[0, t, 0\ldots 0 \right]} \in \left(\cQ_N\right)$ for $t > 0$, we require $a > b$. Now, the quadratic form $q_0(\bbx)$ of \eqref{eq:q0}, for  $\bbx$ of \eqref{eq:qallcond_xt},  takes the form 
 \begin{align}
  q_0(\bbx) = a\left[\trh{\bbxt}\bbM\bbxt + \sqrt{\frac{b}{a}}2\rm{Real}\left(\trh{\bbxt}\bbb z \right)\right] - b\tau > 0,  \label{eq:condonql}
 \end{align}
 where the inequality results after applying  Proposition \ref{prop:infimumcond} and the assumption that $\tau \leq \rm{infimum}\left(\trh{\bbxt}\bbM\bbxt  + 2\rm{Real}\left(\trh{\bbxt}\bbb z\right)\right)$. Using \eqref{eq:condonql}, we see that  $\tr{\left[0, t, 0\ldots 0 \right]} \notin \cQ_N$ for $t > 0$. This completes the proof. 
\end{IEEEproof}
The proof of  P2 is now complete with the following proposition and after combining the relations R1 and R2. 
\begin{proposition}
	$\cQ_N \cap \cN = \emptyset \implies {\rm cov}\left(\cQ_N\right) \cap \cN = \emptyset$. 
\end{proposition}
\begin{IEEEproof}
	The condition $\cQ_N \cap \cN = \emptyset $ implies $q_0(\bbx) \geq 0$ and $q_l(\bbx) = 0, l \geq 1$. From Remark \ref{rem:q1cond} and Proposition \ref{prop:Qn}, we have $q_l(\bbx) = 0$ for $ l \geq 1$ only at $\bbx = \tr{\left[\sqrt{0.5}\tr{\bbxt} \sqrt{0.5}z\right]}$, where $\bbxt$ and $z$ satisfy \eqref{eq:qallcond_xt}. 	At such an $\bbx$, the condition $q_0(\bbx) \geq 0$ implies 
\begin{align}
& 0.5\left[\trh{\bbxt}\bbM\bbxt + 2\rm{Real}\left(\trh{\bbxt}\bbb z \right) - \tau \right] \geq 0 \\
& \implies \tau \leq \trh{\bbxt}\bbM\bbxt + 2\rm{Real}\left(\trh{\bbxt}\bbb z \right), \\ 
& \implies \tau  \leq {\rm infimum}\left(\trh{\bbxt}\bbM\bbxt + 2\rm{Real}\left(\trh{\bbxt}\bbb z \right)\right),
\end{align}		
where the infimum is taken over all values of $\bbxt$ and $z$ satisfying \eqref{eq:qallcond_xt}. Thus, after using Proposition \ref{prop:conichullq}, we have that $\tr{\left[0, 1, 0\ldots 0 \right]} \notin {\rm con}\left(\cQ_N\right)$. This implies that the \emph{origin is boundary point} of ${\rm con}\left(\cQ_N\right)$. A necessary and sufficient condition for  origin to be a boundary point is  existence of a point that does not belong to ${\rm con}\left(\cQ_N\right)$  \cite{dines}. Thus, $\cQ_N \cap \cN = \emptyset \implies {\rm con}\left(\cQ_N\right) \cap \cN = \emptyset  \implies {\rm cov}\left(\cQ_N\right) \cap \cN = \emptyset$. 
\end{IEEEproof}
\bibliographystyle{IEEEtran}
\bibliography{IEEEabrv,references}
\end{document}